\documentclass[notitlepage,twocolumn,prl,longbibliography]{revtex4-2}

\usepackage{amsmath}
\usepackage{changes}
\usepackage{graphicx}
\usepackage[space]{grffile} 
\usepackage{gensymb}
\usepackage{bm}
\usepackage{float}
\usepackage{amssymb,amsfonts,amsmath,amsbsy,bm,t1enc,latexsym}
\DeclareMathOperator*{\argmax}{argmax}
\usepackage{changes}
\usepackage{soul}
\usepackage[super]{nth}
\usepackage[normalem]{ulem}
\usepackage[
bookmarks=true,
colorlinks=true,
linkcolor=blue,
urlcolor=blue,
citecolor=blue,
pdftex,
bookmarks=true,
linktocpage=true, 
hyperindex=true
]{hyperref}
\usepackage{hyperref}
\usepackage{siunitx}
\usepackage[super]{nth}

\begin{document}	
	
\title{Chaotic micro-comb based parallel ranging}
	
	\author{Anton Lukashchuk}
	\author{Johann Riemensberger}
	\author{Aleksandr Tusnin}
	\author{Junqiu Liu}
	\author{Tobias J. Kippenberg}
	\email{tobias.kippenberg@epfl.ch}
	\affiliation{Laboratory of Photonics and Quantum Measurements (LPQM), Swiss Federal Institute of Technology Lausanne (EPFL), CH-1015 Lausanne, Switzerland}
	
	\date{\today}
	
	
	\maketitle




\textbf{
The transition to chaos is ubiquitous in nonlinear systems ranging from fluid dynamics and superconducting circuits to biological organisms \cite{Strogatz2018}. 
Optical systems driven out of equilibrium such as lasers \cite{Haken1975} and supercontinuum generation \cite{Dudley2006} exhibit chaotic states of light with fluctuations of both amplitude and phase and can give rise to Levy statistics, turbulence, and rogue waves \cite{Otsuka2000, Grelu2015}. 
Spatio-temporal chaos also occurs in continuous-wave driven photonic chip based Kerr micro-resonators \cite{Herr2012}, where it is referred to as chaotic modulation instability.
Such modulation instability states have generally been considered impractical for applications, in contrast to their coherent light state counterparts, which include soliton \cite{Herr2014} or dark-pulse states \cite{Xue2015}. 
Here we demonstrate that incoherent and chaotic states of light in an optical microresonator 
can be harnessed to implement unambiguous \cite{Horton1959} and interference-immune \cite{guosui_development_1999} massively parallel coherent laser ranging by using the intrinsic random amplitude and phase modulation of the chaotic comb lines. 
We utilize 40 distinct lines of a microresonator frequency comb operated in the modulation instability regime. 
Each line carries $>$1 GHz noise bandwidth, which greatly surpasses the cavity linewidth \cite{Matsko2013}, and enables to retrieve the distance of objects with cm-scale resolution.
Our approach utilizes one of the most widely accessible microcomb states, and offers - in contrast to dissipative Kerr soliton states - high conversion efficiency, as well as flat optical spectra, and alleviates the need for complex laser initiation routines. Moreover the approach generates wideband signal modulation without requiring any electro-optical modulator or microwave synthesizer. 
Viewed more broadly, similar optical systems capable of chaotic dynamics could be applied to random modulation optical ranging as well as spread spectrum communication \cite{Esman2016} and optical cryptography systems \cite{Chen2004}.
}

The term Random Modulation Continuous Wave (RMCW) denotes a number of microwave and optical techniques for ranging and velocimetry, where random amplitude or phase modulation of a carrier is used to interrogate a target. Conventionally, the modulation is encoded in the form of a random bit sequence or white noise.
A noise-modulated measuring system was first proposed in 1959 \cite{Horton1959}, highlighting unambiguous ranging as an important advantage over periodically modulated continuous-wave (CW) range finders.
It was later realized that a randomly modulated signal occupying wide bandwidth can also be applied in electronic warfare supporting low probability of intercept and effectiveness against electronic counter-countermeasures \cite{guosui_development_1999}. 
Indeed, the security and robust immunity to jamming and interference led to development of spread spectrum communications \cite{Scholtz1977} where a broadband transmitted signal resembles naturally occurring noise.
 
RMCW light detection and ranging (LiDAR) was first demonstrated in 1983 by Takeuchi et al. \cite{takeuchi_random_1983} and like its microwave counterpart \cite{Narayanan1998} relies on cross-correlation between the received signal (or decoded modulation data) and the reference (code) for time delay estimation \cite{carter_coherence_1987}.
RMCW systems, similar to other continuous wave LiDARs, can operate with higher average optical powers than pulsed laser systems and allow for coherent averaging, i.e. the signal-to-noise ratio (SNR) increases in linear fashion with the measurement time. However, in contrast to frequency modulated continuous wave (FMCW) LiDAR \cite{Behroozpour2017,Pierrottet2008}, it does not require stringent conditions on frequency agility and linearity of the lasers that trade-off tuning range (i.e. distance resolution) versus coherence length (i.e. detection range) \cite{Amann1992}.
Furthermore, RMCW is capable of dynamically updating the measurement integration time without the need to adjust the emitted waveform \cite{Ahn2007}.
The random bit sequence or broadband noise can be encoded externally using phase or amplitude modulators \cite{bashkansky_rf_2004} or a noisy optical source such as a semiconductor laser with delayed feedback can be utilized directly \cite{Lin2004}.
Similarly to FMCW, the bandwidth of the source determines the resolution \cite{Axelsson2004} and coherent RMCW, employing heterodyne detection to recover the phase of the signal yields range and velocity information simultaneously. 
In the foreseen epoch of unmanned vehicles the immunity to mutual interference with other LiDARs and ambient light sources makes this advantage of RMCW significant. 
Indeed, LiDAR based on random noise modulation achieves lower false alarm rate and decreased SNR reduction penalty compared to other implementations when subjected to interference of other sensors \cite{Hwang2020}.

\begin{figure*}[!htbp]
	\includegraphics[width=\linewidth]{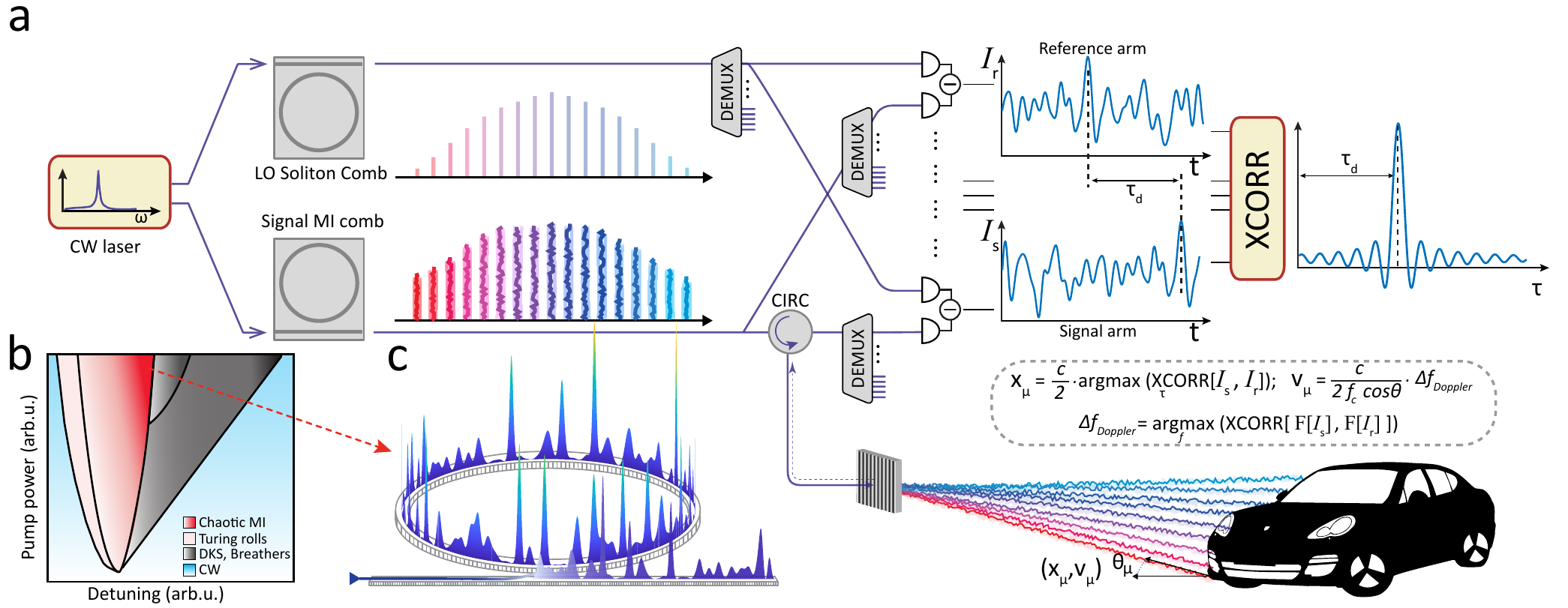}
	\caption{\textbf{Concept of chaotic microcomb coherent LiDAR.}
	a)~Schematic of experimental setup. A single pump laser drives two distinct optical microresonators, generating a chaotic microcomb (signal) and a soliton microcomb (local oscillator, LO) with nearly identical mode spacing.
	The chaotic signal comb is spatially dispersed over the target area using a diffraction grating. Each signal comb tooth $\mu$ represents an independent RMCW channel measuring distance $x_{\mu}$ and velocity $v_{\mu}$ of objects along its beam path. A small portion of the chaotic comb is tapped for the reference measurement.
	The signal, reference and LO comb teeth are filtered with wavelength division multiplexers (DEMUX), superimposed on two balanced photodetectors, and analyzed for time delay estimation. 
	Maximum peak in the cross-correlation trace corresponds to the travel time from which the distance is inferred. The velocity is calculated by measuring Doppler shift via cross-spectrum correlation. 
	b)~Microcomb stability chart illustration. Chaotic modulation instability (MI) area represents the microcomb state of interest. Gradient fill towards top right highlights the preferred region of operation.
	c)~Simulation of chaotic MI waveform in the microresonator.
	}
	\label{fig_concept}
\end{figure*}

Recent works on the application of microresonator frequency combs have largely focused on dissipative Kerr solitons \cite{Grelu2012, Herr2014,Kippenberg2018}, i.e highly coherent low noise states of the driven nonlinear microresontor that have proven to be promising in many applications including metrology \cite{Trocha2018,Suh2018a}, spectroscopy \cite{Suh2016}, coherent communications  \cite{MarinPalomo2017} to name a few. 
Indeed, nearly all applications of photonic chip based microcombs to date have used dissipative coherent structures of light with the exception of optical coherence tomography \cite{Ji2019} where the added intensity noise of the MI state is shown to be detrimental \cite{Marchand2021}.

Here we focus on chaotic frequency combs \cite{Matsko2013, Godey2014} possessing complex nonlinear dynamics that we leverage to implement a novel type of parallel LiDAR. We emphasize that in contrast to previous applications of frequency combs, the technology described in this letter uniquely requires the use of incoherent microcomb states

Combining conventional coherent RMCW techniques and the chaotic nature of modulation instability (MI) microresonator frequency combs, we demonstrate a modulator-free massively parallel RMCW LiDAR that provides unambiguous distance and velocity information, attains $\sim$~15~cm resolution enabled by its characteristic $\sim$~1~GHz noise bandwidth. 
Moreover MI microcombs are power efficient and provide a flat-top optical spectrum and support both parallel direct detection for distance and parallel coherent detection for simultaneous distance and velocity measurement with a suitable local oscillator. 
Lastly, individual comb teeth can be spatially dispersed using low-cost diffractive elements \cite{Riemensberger2020} for rapid 3D imaging.

\begin{figure*}[!htbp]
	\includegraphics[width=\linewidth]{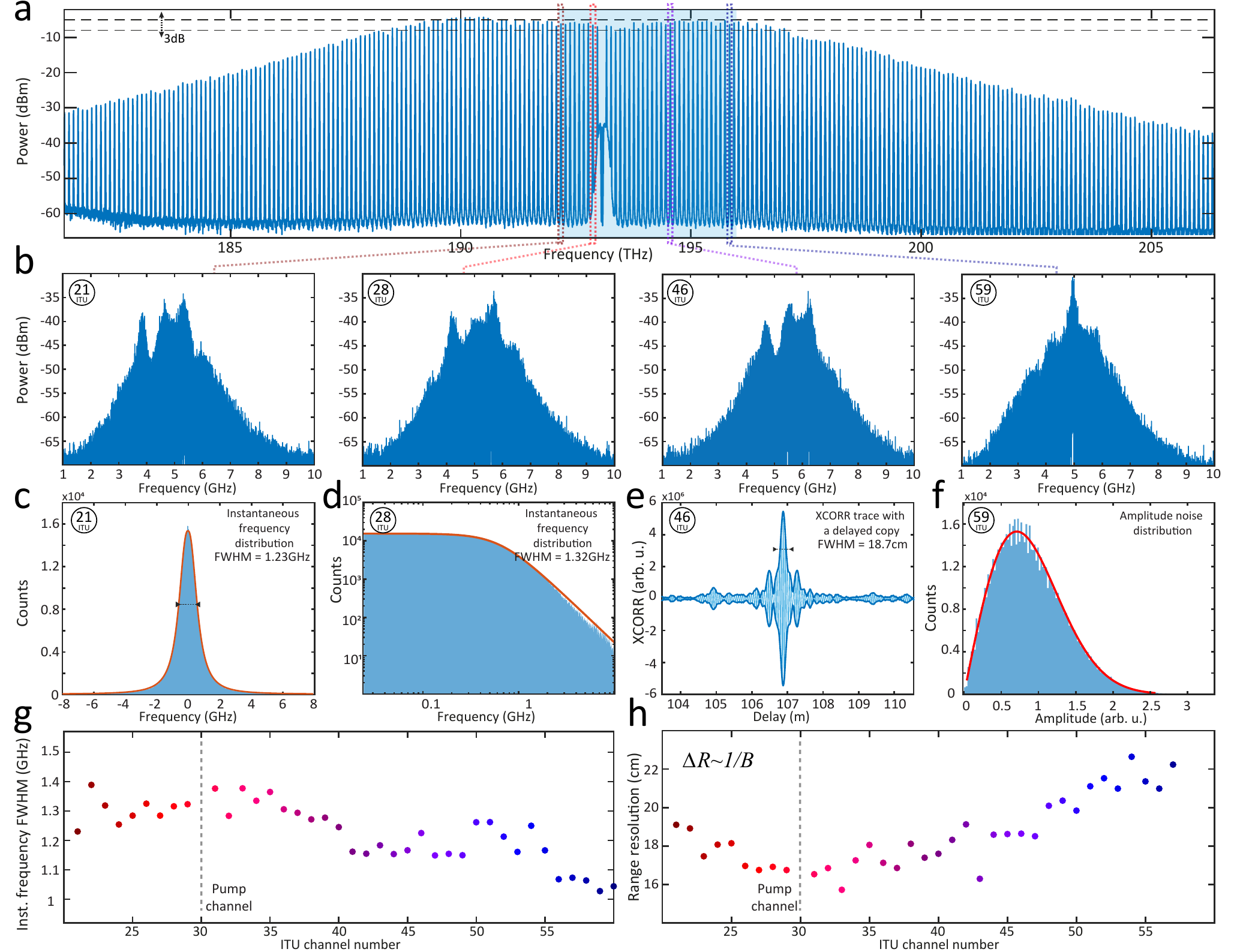}
	\caption{\textbf{Chaotic modulation instability noise properties.}
	a)~Optical spectra of MI comb coupled out of the chip. The residual pump at 193~THz is filtered out. The blue shaded region covers the ITU DWDM channels 20-60 employed in our experiments. Two black dashed lines indicate 3~dB power variation and highlight $\approx$80 comb lines having small power variation.
	b)~Heterodyne beatnote spectra for ITU channels 21, 28, 46, 59 correspondingly.
	c,d)~Distribution of the instantaneous frequency with subtracted mean for ITU channel 21, 28 fitted by Cauchy distribution with $\sim$~2.4 power law.
	e)~Cross-correlation of the signal and the reference for channel 46.   
	f)~Amplitude distribution for ITU channel 59 fitted by Rice distribution.
	g)~FWHM of instantaneous frequency as a metrics characterizing frequency noise bandwidth vs ITU channel number.    
	h)~Channel-dependent distance resolution retrieved from cross-correlation. The distance resolution is consistent with the inverse proportionality to the characteristic noise bandwidth.
	}
	\label{fig_MI}
\end{figure*}


\subsection{Concept of chaotic modulation instability coherent LiDAR}


The principle of random modulation microcomb coherent LiDAR is illustrated in Fig. \ref{fig_concept}a.  
Its operation follows prior works on RMCW systems \cite{Narayanan1998,Hwang2020} but exploits multiple independent optical carriers for line illumination and facilitates parallel detection. 
We utilize a CW laser that pumps two integrated Si$_{3}$N$_{4}$ microring resonators \cite{Liu2018}  with similar mode spacing of 99~GHz. 
Chaotic MI (cf.~Fig.~\ref{fig_concept}b,c) generated in one resonator serves as an array of signals and a dissipative Kerr soliton generated in the second resonator serves as an array of local oscillators (LOs). 
Part of the signal is dispersed by an optical grating to scan the scene. 
The reflected light is demultiplexed and beat with the corresponding LO to produce the signal current. 
Another part of the signal generates the reference current with a second part of the LO. The currents are digitized and cross-correlated to infer distance and velocity of the target. 
To individually access the optical lines three units of commercial dense wavelength division (DWDM) demultiplexers covering the range between 192-196~THz (20-60 ITU channels) are used. 
The system requires 2$\times$N balanced photodiodes and analog-to-digital converters where N is the number of channels detected in parallel. 
Unlike conventional dual-comb implementations where signal and LO combs are detected on a single receiver \cite{Coddington2016, Lukashchuk2021}, high noise bandwidth of MI precludes this implementation. 
The distance information can be inferred from the peak location estimate of the correlation trace \cite{carter_coherence_1987}. 
Cross-spectrum correlation of the signal and reference currents results in a Doppler shift. 
The two combs operate in a free-running regime and do not require any stabilization. 
Unlike FMCW-soliton based implementations of massively parallel coherent laser ranging \cite{Riemensberger2020,Lukashchuk2021}, this approach does not depend on either linearization via phase-locked loop, predistortion of the pump waveform, nor channel calibration. MI  state is obtained deterministically, does not require complex tuning \cite{Guo2017} and is thermally self-locked.

\subsection{Chaotic modulation instability characterization} 

A key advantage of the present implementation of massively parallel coherent laser ranging is the use of chaotic microcombs to directly generate a large number of broadband optical noise states in the optical microresonator with high power conversion efficiency. The optical spectrum of the chaotic microcomb after filtering out the residual pump laser is depicted in Fig.~\ref{fig_MI}a. 

It reaches high power per comb line of up to \SI{-5}{\decibel\meter} with a flat-top spectral shape spanning $\sim$~\SI{8}{\tera\hertz} in \SI{3}{\decibel}, which significantly exceeds the nonlinear conversion performance of soliton microcomb generation in the same resonator, which is depicted in the SI for comparison. 


We attain >20$\%$ power transfer efficiency while typical microresonator soliton state reaches less <1$\%$ \cite{Wang2016} and is theoretically limited to $\sim$~2$\%$ \cite{Bao2014}.
We have deliberately chosen an overcoupled resonance with a total linewidth of 180~MHz (cf. microresonator characterization in SI) to increase the power per comb line and to attain chaotic MI states with higher noise bandwidth (cf. SI).

The dynamics of the Kerr combs in monostable regime (i.e. on the blue-detuned side of the resonance) was studied in \cite{Herr2012}. Furthermore, it was shown that gradual increase in the pump power leads to transition from the modulation instability state to the spatiotemporal chaos \cite{Coulibaly2019}. The resulting noise bandwidth of the waveform increases with the pump power which can be evidenced from the narrowing auto-correlation trace of the temporal intracavity field profile. 
In our experiments, we slowly tune the pump towards the resonance keeping the input power constant. 
As we show experimentally and computationally (see SI), the noise bandwidth of every single chaotic MI comb line increases when the pump is further tuned in to the resonance resulting in narrower distance auto-correlation traces and smaller distance resolution.  
It also leads to the largest bandwidth, spectral flatness and power conversion efficiency. 
Furthermore, RF noise bandwidth of the chaotic MI state comb lines depends on the microresonator Kerr frequency shift that is maximized for the overcoupled resonances. Counter-intuitively, the characteristic noise bandwidth of a given comb line can reach more than 3~GHz with increasing pump power significantly exceeding the cavity linewidth \cite{Matsko2013}.

We utilize a single sideband modulator prior to soliton generation to offset it by 5~GHz from the signal MI.
It enables us to unambiguously measure Doppler shifts without employing phase diversity detection \cite{Blackmore2019}, allows AC coupling of the photoreceviers (our RF amplifier has highpass cut-off at 550~MHz), and unambiguously separates phase and amplitude noise. 
Figure \ref{fig_MI}b illustrates heterodyne spectra of MI combs with corresponding LOs. 
Due to the irregular spectra of the individual chaotic MI comb lines, we analyze the instantaneous frequency \cite{Boashash1992} using a Hilbert transform of the beat with the coherent soliton comb line LO to estimate the effective frequency noise bandwidth $B_\mu$  (cf.~Fig \ref{fig_MI}c,d).
We find that instantaneous frequency distribution is best fitted with a Cauchy distribution of power $\sim$~2.4.
Similar to conventional RMCW or FMCW LiDAR implementations, the distance resolution is directly related to the total noise bandwidth (RMCW) or chirp bandwidth (FMCW) as $\Delta R_\mu = c/2B_\mu$.
The cross-correlation between the digitized beat notes of the signal and reference arms, where the signal is delayed in \SI{107}{\meter} of optical path length is depicted in Fig.~\ref{fig_MI}e. 
The line-by-line dependence of the effective frequency noise bandwidth and the FWHM of the cross-correlation, which determines the smallest distance to distinguish between two semitransparent objects along the beam path, is depicted in Fig.~\ref{fig_MI}g,h. 
As expected the two quantities are inversely proportional.
 
The dynamics in out-of-equilibrium nonlinear optical systems can exhibit non-Gaussian behavior, and give rise to Levy Statistics, i.e. heavy-tailed distributions, which lead to the appearance of rogue waves \cite{Akhmediev2009, Dudley2014}.
Such rare events were observed in high power regime for the blue-detuned Kerr cavity \cite{Coulibaly2019}.
In a practical system, rogue waves might lead to power spikes in the amplifiers and detectors. 
We did not observe any penalty on LiDAR operation in our lab experiments and line-by-line heterodyne detection with a strong local oscillator does mitigate the effects of rogue waves on the photoreceivers. 
We find that the amplitude of any particular comb line is well described by Rice's distribution \cite{Rice1945} as expected for Gaussian white noise in both field quadratures and its effective noise bandwidth corresponds to the instantaneous frequency bandwidth (see SI).
That means that each chaotic MI channel could also readily be applied for direct detection RMCW LiDAR systems at no resolution penalty and without the soliton LO. 


\begin{figure*}[!htbp]
	\includegraphics[width=\linewidth]{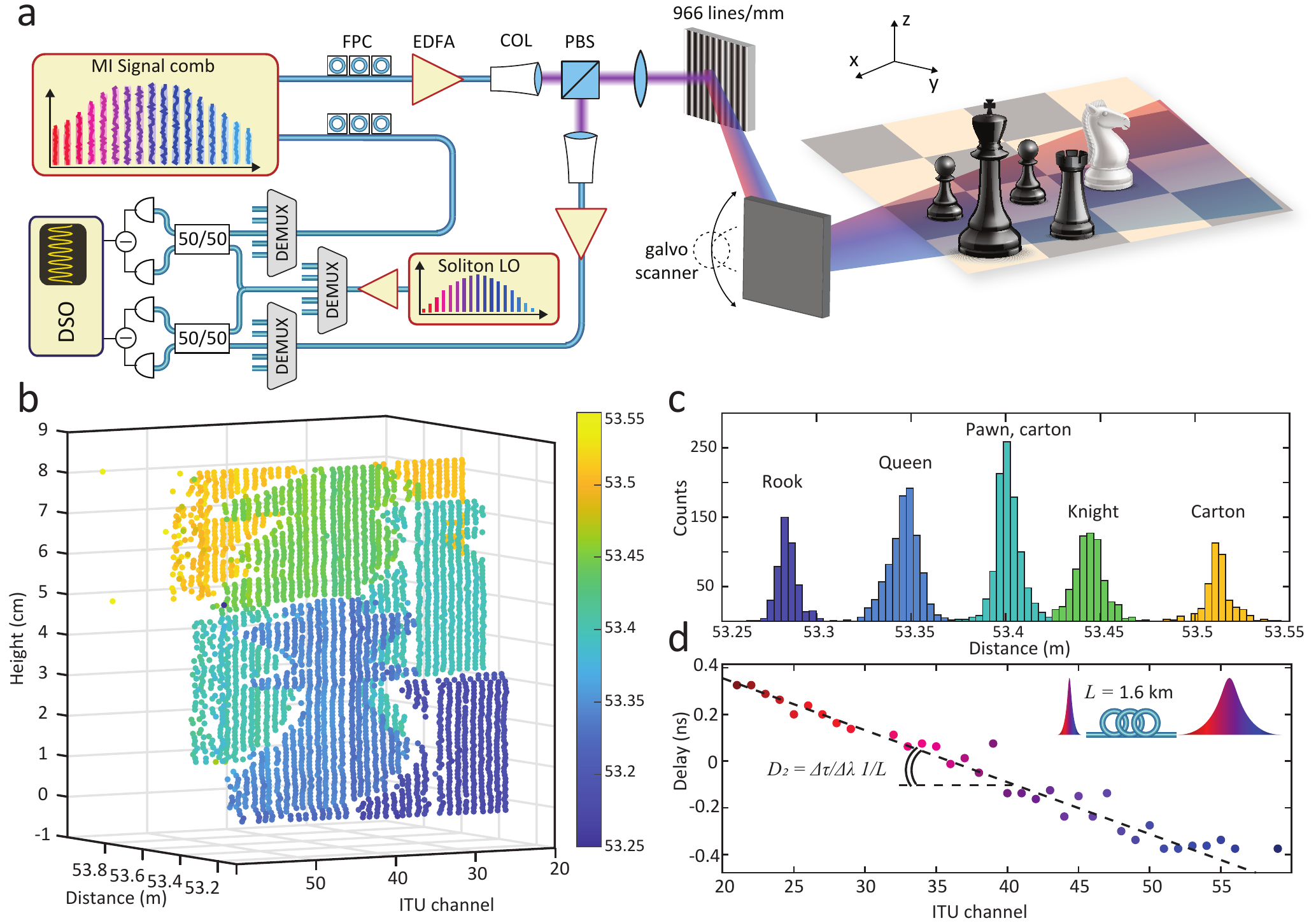}
	\caption{\textbf{Chaotic modulation instability micro-comb based 3D imaging.} 
	a)~Experimental setup. The amplified signal comb is dispersed in free-space by a 966 lines/mm transmission grating and vertical scanning is provided by a mirror galvanometer. FPC: Fiber polarization controller; COL: collimator; PBS: Polarizing beamsplitter. 
	b)~Point cloud of the four chess figures and a carton background obtained during a scan (40 $\times$ 100 points) of the mirror galvanometer. 
	c)~Detection histogram of panel b). The results demonstrate high relative accuracy of the measurements. 
	d)~Chromatic dispersion (D$_2$) measurement of 1.6~km Corning fiber patch cord. The inferred dispersion of 17.2~ps/nm/km shows high channel-to-channel accuracy.
	}
	\label{fig_ranging}
\end{figure*}

\subsection{Ranging}

We perform proof-of-principle ranging experiments demonstrating potential of chaotic frequency comb based RMCW LiDAR.
Imaging experimental setup is depicted in Fig. \ref{fig_ranging}a. The signal MI comb is amplified up to \SI{2}{\watt} and dispersed in free-space by a 966 lines/mm grating. 
The beam is scanned in the vertical direction by a galvo mirror.
Four chess figures (Rook, Queen, Pawn, King) \ref{fig_ranging}b are placed approximately $\sim$~1~m in front of the beam-splitter and galvo mirror.
Major part of the distance delay (105~m) comes from the fiber path difference between reference and signal arms.
The measurement time of one pixel equates to \SI{10}{\micro\second} and the full pixel histogram of the scene is depicted in Fig.~\ref{fig_ranging}c. 
The detection is performed sequentially with 40 operational optical channels. 
The histogram of the pixel detections demonstrates relative accuracy of the ranging.
Figure (SI) depicts a second cross-correlation of the same data, but for integration time 1~$\mu$s. 
As expected, the longer 10~$\mu$s acquisition allows for better detection of faint signals such as object edges and low SNR sensing, which is easily seen in the ranging channels adjacent to the pump laser, which are corrupted by excess ASE or channels close to the edges of the signal amplification bandwidth.

To verify distance measurement accuracy of the chaotic frequency comb for different comb lines, we have performed a fiber chromatic dispersion measurement. 
The specified dispersion <18~ps/nm/km of single-mode optical fiber \cite{Corning} corresponds to $\sim$~0.7~ns delay or $\sim$~25~cm difference in measured distance between 192 and 196~THz. 
The fitted slope for 35 channels on a delay-wavelength axis (cf.~Fig.\ref{fig_ranging}d) yields a value of 17.2~ps/nm/km. 
This shows that channel-to-channel accurate and precise distance measurements are possible for every channel without any need for prior system calibration and linearization. 

\begin{figure*}[!htbp] 
	\includegraphics[width=\linewidth]{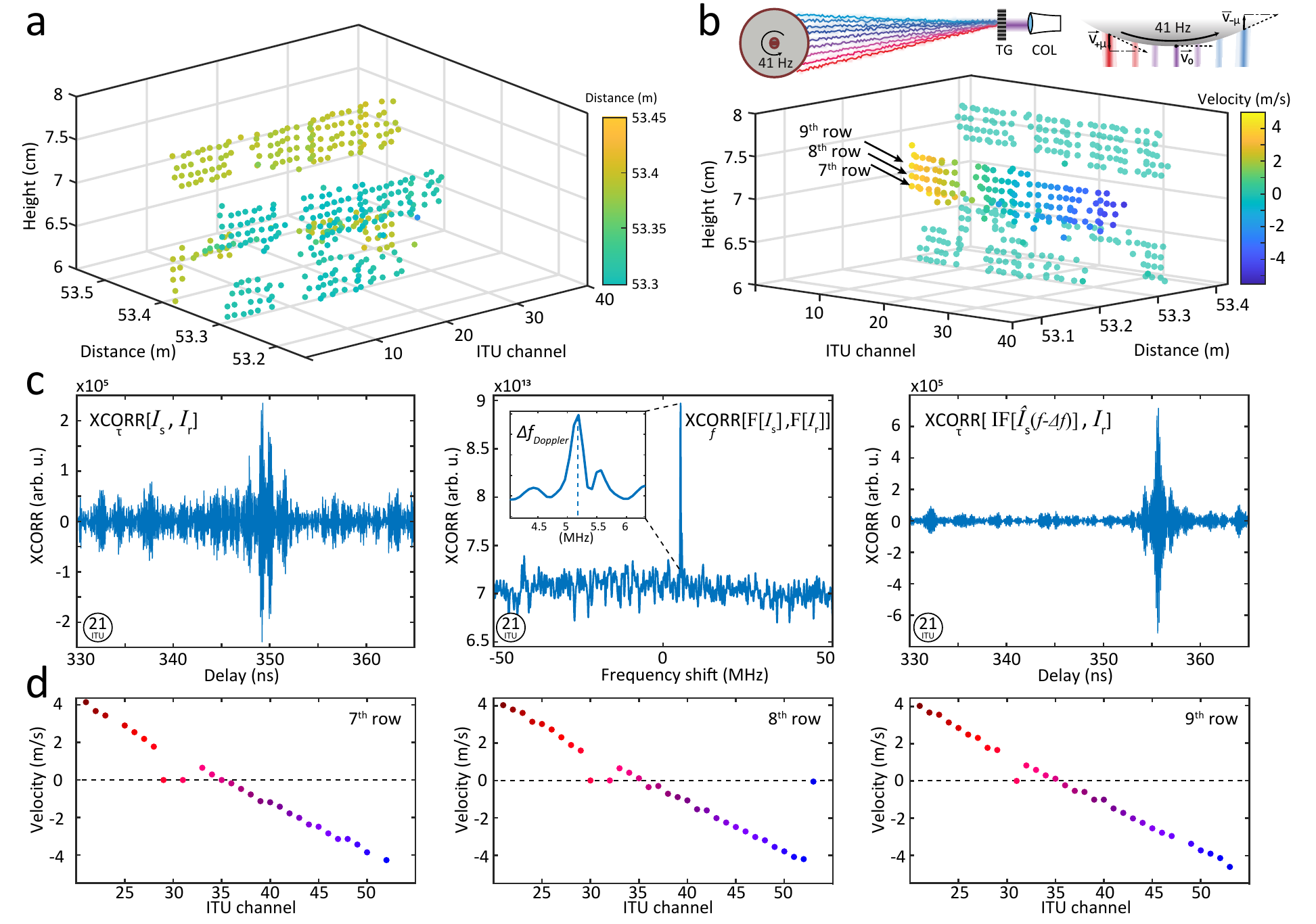}
	\caption{\textbf{Chaotic modulation instability micro-comb based 3D velocimetry.}
	a)~Point cloud of a flywheel. Colormap represents distance information. 
	b)~Point cloud of a flywheel with a different perspective. Colormap represents radial velocity information. 
	c)~Distance and velocity inference procedure. (Left) Direct cross-correlation of signal and reference currents. (Middle) Cross-spectrum correlation highlighting the Doppler shift. Inset corresponds to the beatnote zoom-in with a Fourier transform limited resolution of 100~kHz. (Right) Cross-correlation of frequency downshifted signal and reference currents resulting in a true time delay estimation.  
	d)~Velocity profile cross-sections of the flywheel obtained from panel b).
	}
	\label{fig_velo}
\end{figure*}

\subsection{Velocimetry}

In the second set of experiments, we perform distance and velocity imaging of  a small aluminum flywheel that is covered with a lambertian reflector (cf.~Fig. \ref{fig_velo}a,b).
The motor rotation frequency is \SI{41}{\hertz} and the radius of the wheel is \SI{2}{\centi\metre}.
Direct cross-correlation of the signal and reference heterodyne beat notes (Fig. \ref{fig_velo}c(left)) results in a obscured trace with decreased SNR and shifted time delay.
The problem of cross-correlation degradation in presence of Doppler shift is known from analog noise radar, where it imposes an upper limit on number of samples or the measurement integration time \cite{Axelsson2004}.
However, this issue can be easily solved in digital domain via Doppler correction \cite{Blackmore2019}. 
The Doppler shift is determined by taking a cross-spectrum correlation between signal and reference currents. The peak at $\sim$~\SI{5.2}{\mega\hertz} on Fig.~\ref{fig_velo}c(middle) corresponds to a Doppler shift of ITU channel 21 detecting a pixel moving at $\sim$~4~m/s towards the beam. 
The signal current needs to be downshifted by a Doppler frequency to result in a true delay estimation when cross-correlated with a reference (Fig. \ref{fig_velo}c(right)). 
The SNR is also substantially improved compared to correlation with the uncorrected trace. 
The velocity resolution depends on interrogation time and equates to the Fourier transform limited linewidth, which results in \SI{16}{\centi\meter/\second} velocity resolution for \SI{10}{\micro\second} pixel integration time. 
Figure \ref{fig_velo}d presents the measured horizontal velocity profile of the flywheel.
The $\pm 1$ channels relative to the pump give false detection due to the excess of ASE noise in the soliton LO spectrum. 
Small variations of the velocity are attributed to mechanical vibrations during the line-by-line measurement. 
The acoustic spectrum of the flywheel is depicted in the SI.


\subsection{Summary}

In summary, we have demonstrated a novel architecture for massively parallel chaotic microcomb coherent LiDAR. 
Unlike other applications of microresonator frequency combs, which are generally based on dissipative Kerr solitons, we focus on  incoherent MI combs. 
We actively employ the complex dynamics in a driven Kerr nonlinear photonic microresonator that enables chaotic states of light. Spatio-temporal chaos replaces external modulators that imprint a broadband noise on each of the comb lines.

In our experiments, we utilize 40 comb lines with wideband GHz amplitude and phase modulation, which allows for interference free and unambiguous implementation of massively parallel coherent 3D imaging. 
In contrast to our earlier work on parallel FMCW LiDAR \cite{Riemensberger2020,Lukashchuk2021}, the desired operational state of broadband chaotic MI has much greater optical efficiency, is reached deterministically, does not require soliton switching and it is not necessary to sweep the laser once the desired operation state is reached.  
Another big advantage is that depending on the operating conditions the measurement time can be dynamically updated a posteriori on a per channel basis to increase or decrease measurement sensitivity (range) and ensure the detection of faint objects or reduce computational complexity.
Full heterogeneous integration of InP/Si semiconductor lasers and ultralow-loss silicon nitride microresonators for DKS/MI generation~\cite{Xiang2021} is straightforward and feasible. 
Semiconductor optical amplifiers co-integrated on the silicon substrate \cite{de2020heterogeneous,Vallaitis2010} could potentially replace Er-doped fiber amplifiers. Nonlinear impairment of the signal does not penalize the ranging performance as long as the reference arm is split from the signal arm after the amplification step and the noisy output waveform is the same in both arms. 
The drawback of the architecture is that it requires 2$\times$N balanced photodetectors and three wavelength division demultiplexer arrays. 
However, integration of demultiplexers is straightforward via arrayed waveguide gratings and chipscale integration of large banks of balanced photodetectors has been recently demonstrated \cite{Rogers2020}. 
The sampling requirement of our system is on the order of 2~GS/s and compatible with silicon photoreceivers \cite{Doerr2015}.
The random modulation and cross-correlation based ranging approach does not mandate the use of frequency-agile low noise lasers that would trade of tunability versus linewidth \cite{Amann1992}, nor a high degree of coherence between the signal and local oscillator. 
Waveform monitoring, linearization and predistortion are not required.
Moreover, our system is able to parallelize detection in spectral scanning architectures for 3D imaging \cite{Jiang2020,Baraja2020}, which allows to greatly increase the photon budget without raising eye safety concerns.

Lastly, we would like to highlight further potential applications of MI microcombs where the notion of chaos is actively utilized. 
First, optical spread spectrum communication would be possible if the data is encoded on noisy MI carriers and securely transmitted over a wide bandwidth \cite{Esman2016}.
Spatiotemporal chaos synchronization \cite{Pecora1990,Hu1997} of MI states in remote optical microresonators could directly benefit the encoding and decoding of information on multiple carriers in such applications.
Finally, cryptography based algorithms share a lot with chaotic systems \cite{Kocarev2001} and chaotic maps have already been successfully applied to image encryption and decryption \cite{Chen2004} for example.

\section{Acknowledgements}
This material is based upon work supported by the Air Force Office of Scientific Research under Award No. FA9550-19-1-0250. The work was moreover supported by the Swiss National Science Foundation (SNF) under contract No. 192293.
A.L. acknowledges support from the European Space Technology Centre with ESA Contract No. 4000133568/20/NL/MH/hm.
J.R. acknowledges support from the Swiss National Science Foundation under grant no. 201923 (Ambizione). 
A.T. acknowledges support from the European Union’s Horizon 2020 research and innovation programme under the Marie Skłodowska-Curie grant agreement No. 812818 (MICROCOMB). The Si$_3$N$_4$ samples were fabricated in the EPFL Center of MicroNanoTechnology (CMi). 
Authors thank Alexey Tikan for valuable comments and discussions.
Chess model vector graphics in Fig. 3 was taken from Freepik.com.

\section{Data Availability Statement}
All data, figures and analysis code will be published on \texttt{Zenodo} upon publication of the work.

\section{Author contributions}
A.L. and J.R. conceived and developed the idea. A.L. conducted the various experiments and analyzed the data. A.L. designed the samples. J.L. fabricated the samples. J.R. characterized the samples.  A.T. performed numerical simulations. T.J.K. supervised the work. All authors contributed to the manuscript. 

\bibliographystyle{apsrev4-1}
\bibliography{citations}

\section{Methods}

\subsection{Sample fabrication}
\noindent
Integrated Si$_{3}$N$_{4}$ microresonators are fabricated with the photonic damascene process~\cite{Pfeiffer2018d}. Features are defined using deep-ultraviolet (DUV) stepper lithography~\cite{Liu2018} and reactive ion etching and silica preform reflow~\cite{Pfeiffer2018} prior to deposition reduces scattering losses. 
The waveguide width, height and radius are \SI{1.5}{\micro\meter}, \SI{0.82}{\micro\meter} and \SI{228.43}{\micro\meter} correspondingly.
The positions of the resonance frequencies close to the pumped resonance are expressed with the series $\omega_{\mu} = \omega_0 + \sum_{i \geq 1} D_i \mu_i / i!$. 
The dispersion, loss, and bus waveguide coupling of the microresonator was characterized via optical frequency comb assisted diode laser spectroscopy \cite{Pascal2009}. 
The calibrated laser scan is shown on Extended Data Fig. \ref{fig_SI_res_characterization}a, where the free spectral range equates to $D_{1}/2\pi=$ \SI{98.9}{\giga\hertz}. 
The integrated dispersion $D_\textrm{int} = \omega_{\mu}
 - \omega_0 - D_1\mu$ is depicted in extended data Fig. \ref{fig_SI_res_characterization}b and a polynomial fit reveals anomalous second order dispersion of $D_{2}/2\pi=$ \SI{1.5}{\mega\hertz}. 
The LO comb  microresonator has similar design cross-section and radius, which results in a FSR offset as low as 40~MHz.
The MI microresonator is operated in the strongly overcoupled regime in order to maximize comb output power. (Extended Data Fig. \ref{fig_SI_res_characterization}c) with intrinsic loss rate $\kappa_0/2\pi=$ \SI{17}{\mega\hertz} and bus waveguide coupling rate $\kappa_\mathrm{ex}/2\pi=$ \SI{160}{\mega\hertz}.


\subsection{MI and soliton microcomb generation}
\noindent
We use an external cavity diode laser (ECDL) for our proof-of-principle. 
It is split into two arms, where one pumps the signal resonator and the other is offset by $\sim$~5~GHz by a dual Mach-Zehnder modulator biased for single-sideband regime to pump the LO microresonator.
Each arm is amplified, and coupled into a photonic chip using lensed fibers.
Optical power incident on either of the chips is $\sim$~2~W. 
Manual tuning with the ECDL piezo is used to tune into both resonances at the same time.
The LO soliton comb is tuned into a single soliton state using well established soltion state switching techniques\cite{Guo2017}, while MI comb is tuned to the furthest point before the collapse into the soliton state.
During the soliton switching process, we monitor the laser-cavity detunings via the phase modulation response technique introduced in \cite{Guo2017}. 
Temperatures of the samples are adjusted during the dual tuning process, if necessary using a thermal actuator (Peltier element). 
In this way, MI and soliton states in both resonators can be obtained routinely and quickly.

The spectra of MI and two-soliton states obtained in the MI microresonator are depicted in extended data Fig.~\ref{fig_SI_OSA} and show that the power per comb line is up to \SI{10}{\decibel} higher in the MI state than in the two-soliton state. 
Here we choose to plot a two-soliton state with near perfect symmetry of the solitons for comparison due to the difficulty to obtain single soliton states in the strongly overcoupled MI microresonator. 
The strongest comb lines of the two-soliton state are \SI{6}{\decibel} stronger than in the single soliton state \cite{Herr2014}.

\subsection{Post-processing}
\noindent

In RMCW ranging, distance to the target object $d$  is inferred from the time delay estimation of cross-correlation between signal and reference currents:
\begin{equation}
\mathrm{d} =  \frac{c}{2}\mathrm{\argmax_\tau}(\mathrm{XCORR}[I_s,I_r]).
\label{eq:pp_dist}
\end{equation}
However, if the object is moving, the signal current experiences a Doppler shift and straighforward cross-correlation does not retrieve the true time delay. 
As described in the Velocimetry section of the main manuscript, we calculate first the Doppler shift by cross-spectrum correlation and peak estimation \cite{Blackmore2019}. 
\begin{equation}
\Delta f_{\mathrm{D}}  =  \mathrm{\argmax_\mathit{f}} (\mathrm{XCORR}[\mathcal{F}[I_s],\mathcal{F}[I_r]]),
\label{eq:pp_doppler_shift}
\end{equation}
where $\mathcal{F}[\cdot]$ and $\mathcal{F}^{-1}[\cdot]$ stand for the Fourier and the inverse Fourier transforms, $\Delta f_{\mathrm{D}}$ is a Doppler shift, $f_c$ is an optical carrier frequency. The Doppler shift yields the target velocity via
\begin{equation}
	\mathrm{v} =  \frac{c}{2f_c}\Delta f_{D}.
	\label{eq:pp_vel}
\end{equation}
The true time delay is then easily obtained via cross-correlation of the frequency-shifted signal current with the reference current
\begin{equation}
\tilde{I_s}=  \mathcal{F}^{-1}[\mathcal{F}[I_s](f-\Delta f_{\mathrm{D}})],
\label{eq:pp_signal_correct}
\end{equation}
\begin{equation}
\mathrm{d} =  \frac{c}{2}\mathrm{\argmax_\tau}(\mathrm{XCORR}[\tilde{I_s},I_r]).
\label{eq:pp_dist_correct}
\end{equation}
All cross-correlations and Fourier transforms are calculated digitally in batch processing via MATLAB. However, we note that analog processing of cross-correlations would also be feasible \cite{Zarifi2019}. 

Next, we estimate the computational complexity of massively parallel RMCW optical ranging. Straightforward computation of the cross-correlation yields $O$(N$^2$) with the number of samples N. Recalling that convolution operation can be calculated by the means of the Fourier transforms, i.e. the inverse Fourier transform of the multiplication of the Fourier transforms, the complexity can be estimated as $\approx$3$\times$4N*log$_2$(N) \cite{Frigo1998}. However, taking into account prior information from the preceding step or a neighboring pixel and restricting ourselves to a limited search range around the previous value, we assume that the cross correlation complexity can be estimated as $O$(M*N) if calculated directly, where M$\ll$N and potentially M$<$3$\times$4*log$_2$(N).

\subsection{Numerical simulations}
To verify the experimental results, we perform numerical simulations of the Lugiato-Lefever equation using two distinct numerical schemes: pseudo-spectral 1st order Split-Step method~\cite{Agrawal2000Nonlinear} and the step-adaptative Dormand-Prince Runge-Kutta method of Order 8(5,3)~\cite{press2007numerical} with the second-order finite difference scheme for the dispersion operator. 
Laser power and cavity parameters are chosen to match the experimental settings and results. 
Similar to the experiment, we fix the pump power and scan the laser across the cavity resonance from the red to the blue detuned region as shown in Extended Data Fig.~\ref{fig_SI_MI_evo_res_sim}(b). 
Then, we fix the laser-cavity detuning in the unstable region and investigate temporal dynamics of the optical field at several laser-cavity detunings. 
We compute heterodyne beatnote power spectral densities for several chaotic MI comb lines (see Extended Data Fig.~\ref{fig_SI_MI_evo_OSA_sim}) and find them in very good agreement with the experiment shown in Extended Data Fig.~\ref{fig_SI_MI_evo_OSA_exp} for both numerical schemes. 
Before computing the corresponding auto-correlation traces, we add a \SI{5}{\giga\hertz} frequency offset to the signal similar to the experiment. 
The obtained auto-correlation traces shown in Extended Data Fig.~\ref{fig_SI_MI_evo_XCORR_sim} along with range resolution in Extended Data Fig.~\ref{fig_SI_MI_evo_res_sim} show very good quantitative agreement with the experimental results.

\subsection{Numerical investigation of the noise bandwidth}
In previous sections, we have shown that the maximal resolution is attained in the pre-soliton switching zone, which is depicted in Extended Data Fig.~\ref{fig_SI_FWHM_SPM}a as the edge of the MI zone. Tuning further into the bistable region of the tilted resonance will switch the resonator into one of the coherent regimes: DKS or CW; therefore, the preferred operating point should be at the end of the monostable branch. If the internal linewidth of the device ($\kappa_0$) is fixed, the range resolution can be improved by increasing the coupling $\kappa_\mathrm{ex}$. In order to show it, we consider two identical resonators  with different couplings to the bus waveguide: an overcoupled $\kappa_\mathrm{ex}= 9\kappa_0$ and a critically coupled $\kappa_\mathrm{ex}=\kappa_0$ devices. The corresponding tilted resonances for the input power \SI{0.9}{\watt} in Extended Data Fig.~\ref{fig_SI_FWHM_SPM}a show that the overcoupled device has more optical power (i.e. more photons) in the monostable region. Further, we convert the number of photons to Kerr frequency shift and compute noise bandwidth for a set of comb lines in the chaotic state at the end of the MI zone by performing similar numerical simulations as above. In Extended Data Fig.~\ref{fig_SI_FWHM_SPM}b we show the dependence of the noise bandwidth on Kerr frequency shift for both devices with increased pump power \SI{1.0}{\watt}, \SI{1.1}{\watt},...,\SI{2.0}{\watt}. Both devices have the same trend of resolution improvement with the pump power, but the overcoupled resonator shows better perfomance than the critically coupled one. Increasing of $\kappa_\mathrm{ex}$ decreases the loaded quality factor but provides with more effective photon flux into the cavity. In contrast to the soliton generation, where critically coupled devices provide with larger soliton existence range, the overcoupled resonators in the MI regime have more intracavity power and thus provide with better resolution.

\subsection{Direct detection RMCW LiDAR}
We note there are two possible implementations of RMCW. Coherent detection employing extra local oscillator for every comb tooth and direct detection requiring highly sensitive photodiodes.
While we have demonstrated coherent LiDAR based on heterodyne detection, direct detection chaotic microcomb RMCW LiDAR is equally feasible, since MI possess both amplitude and frequency random-like modulation. Exclusion of the local oscillator on the other side requires highly sensitive photodiodes. Baraja is an example of commercial company utilizing pure amplitude molulation CW LiDAR \cite{Baraja2020}. Given widely used spectral scan approach, chaotic microcombs could be readily applied in the similar setting to remove the necessity for the laser scan and introduce parallelism due the multitude of the comb lines.

Extended Data Fig. \ref{fig_SI_DD} depicts fiber delay distance measurement with $\sim$~1~mm level precision and resolution inferred from cross-correlation traces. Resolution in case of pure amplitude modulation corresponds to the resolution of coherent RMCW LiDAR highlighting similar effective noise bandwidth in amplitude and frequency domains. 

\subsection{Comparison with ASE noise LiDAR}
Evidently, one could argue that ASE noise could be used as an optical source for RMCW LiDAR. Indeed ASE noise from erbium doped fiber amplifier (EDFA) is broadband and follows random-like Gaussian statistics. It would have much higher noise bandwidth even after being filtered by DEMUX unit than a chaotic MI comb line resulting in higher resolution. Extended Data Fig. \ref{fig_SI_ASE} demonstrates distance resolution of ASE noise based RMCW LiDAR versus digital band-pass filter bandwidth. 
However, broadband ASE LiDAR would be less power efficient since part of the signal would be filtered optically in DEMUX unit and another part electronically, as it is challenging to have detectors covering whole RF bandwidth of the ASE noise optical channel. 
If spectral scanning is chosen for parallel operation, MI comb channels are well spectrally separated by FSR of the resonator which in combination with diffractive optics determines angular resolution of the system. At the same time, the ASE source would be equally dispersed without well-defined pixels.
We also envision difficulties in velocity inference, since Doppler shift in case of flat top spectrum fully occupying RF bandwidth is harder to detect than irregular localized spectra of MI comb lines.

\subsection{SNR, resolution and precision of coherent RMCW LiDAR}
\noindent
In this section we will discuss the SNR, resolution and precision of RMCW LiDAR from an analytical standpoint.
To begin with, we would like to draw the readers attention to similarities between conventional FMCW and coherent RMCW ranging techniques. 
Indeed, as we show below both achieve a distance resolution of $c/2B$, where $B$ stands for the sweep amplitude for FMCW and characteristic frequency noise bandwidth in case of RMCW. 
Additionally, the equations for measurement precision also share many similarities.
Also we note that both approaches yield similar SNR and reach single-photon sensitivity.
\subsubsection{SNR}
\noindent
The SNR for shot-noise limited coherent detection with unity quantum efficiency is 
\begin{equation}
\mathrm{SNR} =  \frac{\langle I^2 \rangle}{\langle \Delta I^2 \rangle} = \frac{P^{\mathrm{sig}}}{\hbar\omega B_{\mathrm{RF}}}
\label{eq:SNR_FMCW}
\end{equation}
where $B_{\mathrm{RF}}$ stands for the effective noise bandwidth of the Fourier transform window. This level of sensitivity is expected in conventional FMCW approach \cite{Piggott2020}.

The calculation of the SNR for RMCW includes two steps: the coherent detection of signal and reference and their cross-correlation. The latter closely follows the discussion outlined in chapter 8 of \cite{bendat_random_2011}. To simplify our derivation we will consider the RMCW signal to be bandwidth-limited white Gaussian noise with a zero mean.
We will also consider a simple model for signal $x$ and reference $y$ currents after coherent detection step as
\begin{equation}
\begin{split}
\begin{gathered}
x(t) = s(t-\tau_d) + m(t)\\
y(t) = s(t) + n(t)\\
R_{xx}(0) = R_{ss}(0) + R_{mm}(0) = S + N\\
R_{yy}(0) = R_{ss}(0) + R_{nn}(0) = S + M\\
R_{xy}(\tau) = R_{ss}(\tau - \tau_d)
\end{gathered}
\end{split}
\label{eq:currents}
\end{equation}
Where $s(t)$ represents the contribution to the current from the initial random signal, $m(t)$ and $n(t)$ are noise terms that appear after photodetection due to the shot noise, thermal noise and other possible noise sources.
The SNR of the current $y$, for example, reads as $ \mathrm{SNR}_y = \langle s^2 \rangle / \langle m^2 \rangle = R_{ss}(0)/R_{m}(0) = S/M $.
 $\tau_d$ is a delay of the signal current with respect to the reference current. All of these terms are mutually uncorrelated. We did not introduce an attenuation coefficient for the $s(t-\tau_d)$ term to keep the clarity and as it will be evidenced later, the ratio of the amplitudes $S$ and $M,N$ matters rather than the absolute value of $S$. 
To calculate the SNR of the cross-correlation trace between the signal $x$ and the reference $y$ currents we take the maximum of the cross-correlation function squared and divide by its variance $\mathrm{SNR} = R^2_{xy}(\tau_d) / \mathrm{Var}[R_{xy}(\tau)]$. The denominator term in case of bandwidth-limited white Gaussian noise signals $x$, $y$ can be estimated as \cite{bendat_random_2011, Stec2018}
\begin{equation}
\mathrm{Var}[R_{xy}(\tau)] \approx  \frac{1}{2BT} [R_{xx}(0)R_{yy}(0) + R^2_{xy}(\tau)]
\label{eq:xcorr_variance}
\end{equation}
where $B$ is a signal noise bandwidth and $T$ is the time length of $x$ and $y$ and $T \gg \tau_d$. Taking the upper bound for the variance estimate of cross-correlation at $\tau = \tau_d$ we approximate the value of SNR to be
\begin{equation}
\begin{split}
\begin{gathered}
\mathrm{SNR} \approx R^2_{xy}(\tau_d) / \mathrm{Var}[R_{xy}(\tau_d)]\\
= 2BT/[2 + M/S +N/S + (M/S)(N/S)]
\end{gathered}
\end{split}
\label{eq:SNR_xcorr}
\end{equation}
At this point we make two more assumptions. First, the reference current has a much higher SNR than the signal current $(S/M)\gg(S/N)$. It is a reasonable assumption since we can always allocate some amount of signal power straight after generation and send it directly to the detector without any loss, while the target arm will experience free-space optical loss. Second, the SNR of the imaged photocurrent is much less than one $S/N \ll 1$. This is not only a realistic assumption but also a desired property in spread spectrum communications \cite{Esman2016} and military applications supporting optical analogue of electronic counter-countermeasures and low probability of intercept \cite{guosui_development_1999}. In the FMCW approach the SNR of the imaged photocurrent should be more than one since it corresponds to the beatnote of a single frequency (with a width of resolution bandwidth), however in case of RMCW the spectrum is spread out resulting in much lower SNR at the photodetection stage, while it will be boosted taking cross-correlation. Under these assumptions we finally get \cite{Stec2018, Hwang2020}
\begin{equation}
\mathrm{SNR} \approx 2BT \left (\frac{S}{N} \right)
\label{eq:SNR_xcorr_final}
\end{equation}
Effectively, the photocurrent SNR (that is $S/N$) gets multiplied by time-bandwidth product (cf. Extended Data Fig. \ref{fig_SI_SNR_time}). If one uses a Nyquist sampling rate of $f_s = 2B$ the SNR would read as $ \approx f_sT \left (\frac{S}{N} \right)$, where $f_sT$ is simply a number of sampled data points. Further, we will estimate the value of $S/N$, where $S$ stands for the initial random signal current variance and $N$ is  a variance of the photodetection noise, or, equivalently this ratio equals to the ratio of power spectral densities  at $f<B$ of signal and noise currents, considering them to be flat.  We will assume a shot-noise limited heterodyne detection with unity quantum efficiency. To distinguish Signal-to-Noise at different stages we apply the notation 
\begin{equation}
\mathrm{SNR_{corr}} = 2BT \left( \frac{S}{N} \right) =  2BT \ast \mathrm{SNR_{pd}}
\label{eq:SNR_xcorr_and_pd}
\end{equation}
Two random signals one of which is a true signal (target signal) and the other is a random noise have an SNR determined as a ratio of variances of their currents $\mathrm{SNR_{pd}} = \langle \sigma I^2_s \rangle / \langle \sigma I^2_n \rangle$. The variance of the target signal $ \sigma I^2_s  \propto 2P^\mathrm{sig}P^\mathrm{LO}$, while the variance of the noise $ \sigma I^2_n  \propto 2 \hbar\omega P^\mathrm{LO} \Delta \nu_{\mathrm{RBW}}$. The resolution bandwidth $\Delta \nu_{\mathrm{RBW}}$ equals to the sampling rate, since the time-domain variance of the current noise is determined by the number of the shot noise photons collected during the sampling time. The ratio of the variances leads to
\begin{equation}
\mathrm{SNR_{pd}} = \frac{\langle \sigma I^2_s \rangle} {\langle \sigma I^2_n \rangle} = \frac{P^\mathrm{sig}} {\hbar\omega~f_s}
\label{eq:SNR_pd}
\end{equation}

Substituting eq. \ref{eq:SNR_pd} in to eq. \ref{eq:SNR_xcorr_and_pd} and considering sampling rate to be twice as noise bandwidth $f_s = 2B$, we arrive to eq. \ref{eq:SNR_FMCW}, where the resolution bandwidth $B_\mathrm{RF}$ is inverse of the measurement time $T$. 

\subsubsection{Distance measurement resolution}
\noindent
We find it useful to bring analogies between FMCW and RMCW approaches, given they have a lot in common. 
That is why for our next discussion on resolution and precision we will first derive it for the case of FMCW and then compare it to the results of RMCW.
In the context of LiDAR, the term distance resolution usually refers to a longitudinal (or axial) resolution, i.e. the minimum resolvable distance between two reflectors within a single optical  path.
Recall that the value of the beatnote for a given measured distance in FMCW case with triangular ramp waveform modulation follows $f_b = \tau_d \cdot 2B/T = 2R/c \cdot 2B/T$ \cite{Pierrottet2008}.
The minimum separation frequency one can resolve is determined by the measurement time, which often is the same as a chirp period $T$. 
The value of the beatnote is derived by taking the Fourier transform over the up/down ramp with duration $T/2$ resulting in frequency resolution $2/T$. 
Then, the distance resolution reads as
\begin{equation}
\Delta R = \frac{cT} {4B} \Delta f_b = \frac{c}{2B}
\label{eq:_dist_res_FMCW}
\end{equation}

In case of RMCW with flat-top power spectrum with finite bandwidth, the resolution can be inferred directly from the auto-correlation function. 
According to the Wiener-Khinchin theorem, the auto-correlation function is the Fourier transform of the power spectral density of the signal. 
The Fourier transform of a rectangular function with the width $B$ is a sinc function ($ \sin(\pi x)/ (\pi x) $) with first zeros at $t =\pm 1/B$.
Following the convention, two sinc functions can be resolved if the maximum of the second sinc corresponds to the position of the minimum of the first sinc function, which leads to the following expression for RMCW ranging resolution:
\begin{equation}
\Delta R = \frac{c \Delta T}{2} = \frac{c}{2B}
\label{eq:_dist_res_RMCW}
\end{equation}
In case of a random modulation with a pulse sequence, one assumes the pulse width $1/B$ as a time resolution unit and gets the same equation as \ref{eq:_dist_res_RMCW} for distance resolution \cite{takeuchi_random_1983, bashkansky_rf_2004}. 

\subsubsection{Distance measurement precision}
\noindent

Mathematically, precision is a standard deviation of the estimated value and the impact of random noise on the estimation.
Many LiDAR papers provide precision results without further quantification on SNR, measurement time, sampling rate or number of sampled points, all of which strongly impact the measurement precision. 
Sometimes it is mentioned that high SNR enables sub-resolution precision or one can see estimates of precision in FMCW case given as $\Delta R/(SNR)$ \cite{baumann_speckle_2014}, where $\Delta R$ stands for resolution.

Montgomery and O$'$Donoghue \cite{montgomery_derivation_1999} studied a least squares fit problem with a sinusoidal signal in presence of random uncorrelated noise. Analytical formula for precision reads as
\begin{equation}
\sigma_f = \sqrt{\frac{6}{N_s}}\frac{1}{\pi T}\frac{1}{\sqrt{\mathrm{SNR}}} = \sqrt{\frac{6}{f_s}}\frac{1}{\pi \sqrt{\mathrm{SNR}}} T^{-3/2}
\label{eq:prec_sin}
\end{equation}
where $N_s$ is a number of sampled points and other definitions are consistent with the previous sections. The same formula was heuristically derived in \cite{dolan_accuracy_2010}. Similar result with a $\sqrt{2}$ difference was obtained examining Cramer-Rao lower bound \cite{rife_single_1974}.
The estimate for the distance precision can be readily achieved substituting eq. \ref{eq:prec_sin} into eq. \ref{eq:_dist_res_FMCW}
\begin{equation}
\begin{split}
\begin{gathered}
\sigma_R = \frac{cT} {4B} \sigma_f = \frac{c}{2B} \sqrt{\frac{6}{N_s}}\frac{1}{2\pi}\frac{1}{\sqrt{\mathrm{SNR}}}\\ 
=  \sqrt{\frac{6}{N_s}}\frac{1}{2\pi}\frac{\Delta R}{\sqrt{\mathrm{SNR}}}
\end{gathered}
\end{split}
\label{eq:prec_FMCW}
\end{equation}
Indeed, the precision improves with a higher SNR, but follows a square root power law. It also improves with a number of samples, which can be reformulated in terms of measurement time, sampling rate or even chirp bandwidth if the latter is linked to the sampling rate.

For RMCW Carter\cite{carter_coherence_1987} gives a rigorous derivation of the Cramer-Rao lower bound on time delay estimation variance for bandwidth-limited white Gaussian signals with uncorrelated white noise:
\begin{equation}
\begin{split}
\begin{gathered}
\sigma_\tau^2 = \frac{3} {8\pi^2} \frac{1+2\mathrm{SNR_{pd}}}{\mathrm{SNR_{pd}}^2} \frac{1}{B^3T}
\end{gathered}
\end{split}
\label{eq:prec_RMCW}
\end{equation}
where $\mathrm{SNR_{pd}}$ denotes the Signal-to-Noise ratio of the signal channel after the detection prior to cross-correlation with a reference channel.

In case of a high $\mathrm{SNR_{pd}}$ the formula can be simplified and the distance precision reads as:
\begin{equation}
\begin{split}
\begin{gathered}
\sigma_R = \frac{c}{2B}\frac{\sqrt{3}}{2\pi} \sqrt{\frac{2}{\mathrm{SNR_{pd}}}} \frac{1}{BT}\\
= \sqrt{\frac{12}{N_s}} \frac{1}{2\pi} \frac{\Delta R}{\sqrt{\mathrm{SNR_{pd}}}}
\end{gathered}
\end{split}
\label{eq:prec_RMCW_high_SNR}
\end{equation}
where we have considered sampling rate to be twice as noise bandwidth giving the term $N_s$ and the formula resembling eq. \ref{eq:prec_FMCW} for FMCW case.

However in case of RMCW LiDAR the returned signal has low SNR, since the signal is spread over the whole bandwidth and noise variance dominates over the signal variance. The low SNR case is discussed in \cite{carter_coherence_1987, quazi_overview_1981}. Assuming $\mathrm{SNR_{pd}} \ll 1$, eq.~\ref{eq:_dist_res_RMCW} for distance can be written:
\begin{equation}
\begin{split}
\begin{gathered}
\sigma_R = \sqrt{\frac{6}{N_s}} \frac{1}{2\pi} \frac{\Delta R}{\mathrm{SNR_{pd}}}
\end{gathered}
\end{split}
\label{eq:prec_RMCW_low_SNR}
\end{equation}
where in comparison to eq. \ref{eq:prec_FMCW} the dependence on electrical current SNR is just inverse but not squared.

%

\newpage
\section{EXTENDED DATA FIGURES} 
\renewcommand{\figurename}{\textbf{Extended Data Fig.}}
\setcounter{figure}{0}

\newpage
\begin{figure*}[!htbp]  
	\includegraphics[width=\linewidth]{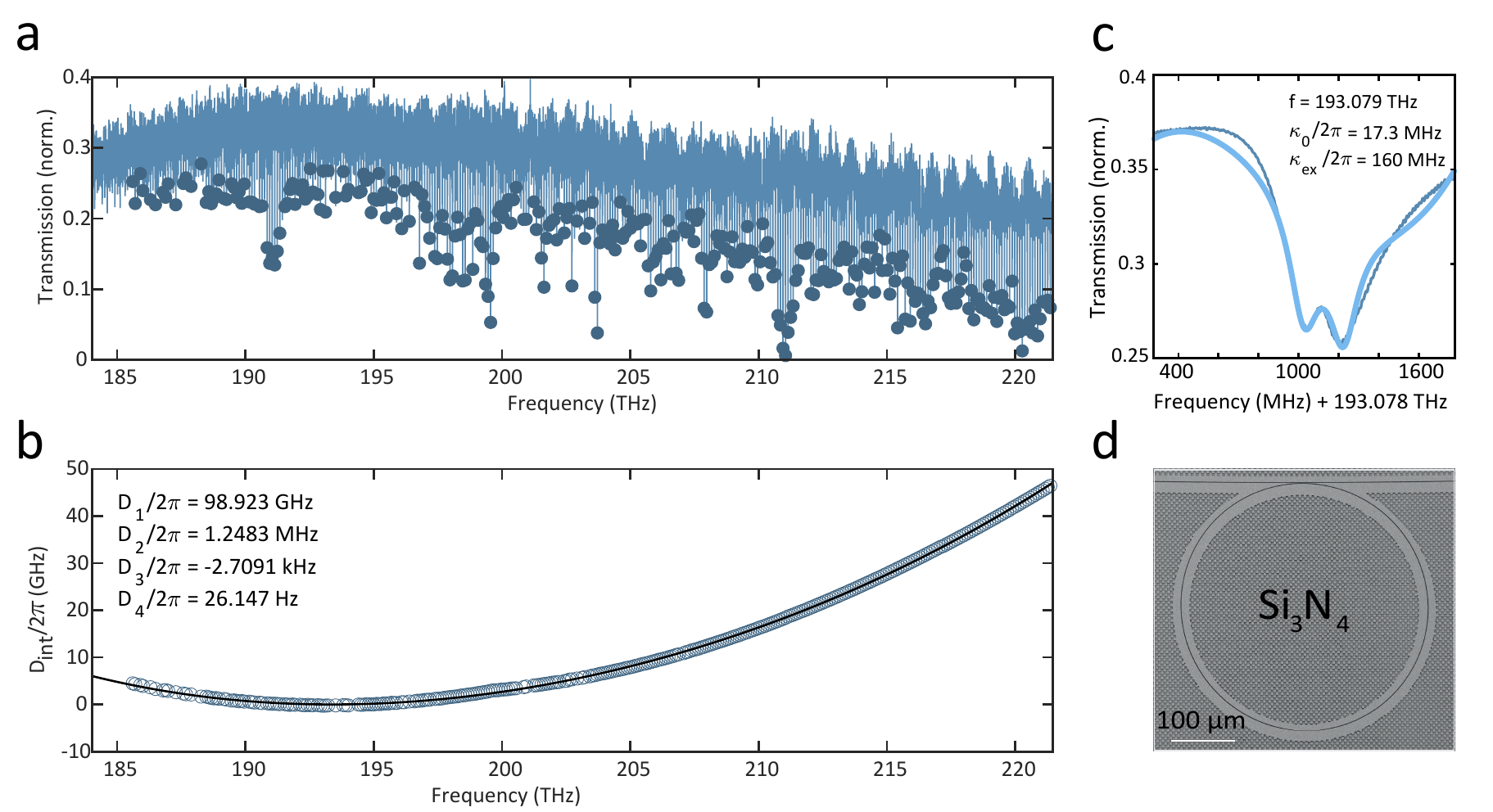}
	\caption{\textbf{MI comb resonator characterization.}
	a)~Transmission scan of the resonator featuring TE mode resonances spaced $\sim$~99~GHz apart.
	b)~Integrated dispersion of TE mode.
	c)~Zoom in of the resonance utilized (pumped). The resonance is highly overcoupled.
	d)~Electron microscope picture of 228.43~$\mu$m Si$_3$N$_4$ microring resonator.
	}
	\label{fig_SI_res_characterization}
\end{figure*}

\newpage
\begin{figure*}[!htbp] 
	\includegraphics[width=\linewidth]{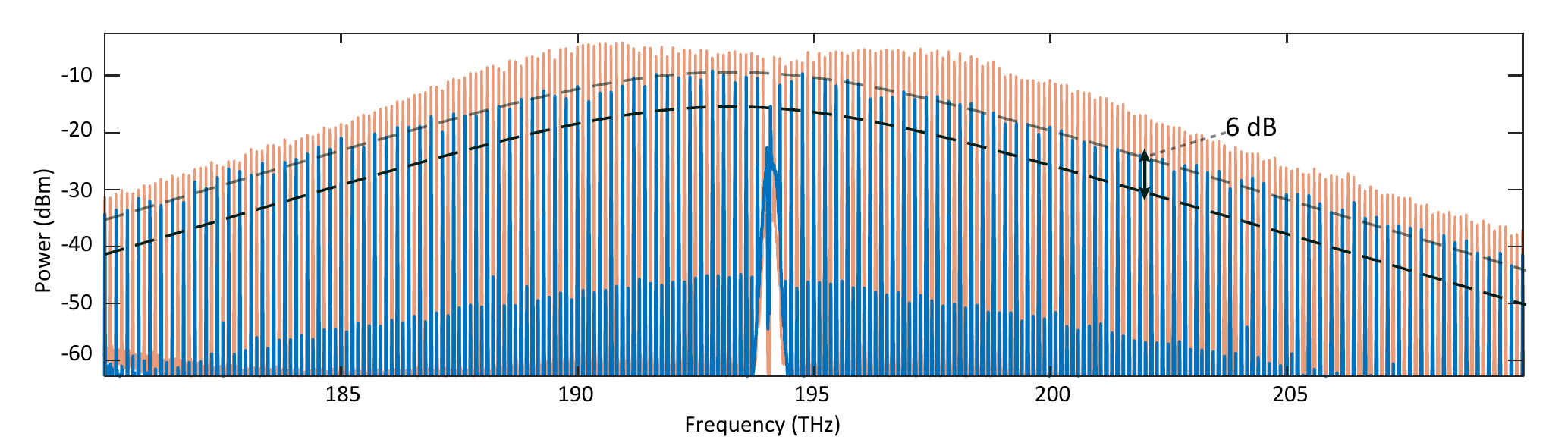}
	\caption{\textbf{Optical spectra of MI and soliton state}
	MI (red) and two-soliton state (blue) spectra obtained in the same resonance and the same pump power. Gray dashed line is a two-soliton sech fit, while black dashed line shows how a single soliton state would look like. The reader is referred to differences in power levels between soliton and MI.
	}
	\label{fig_SI_OSA}
\end{figure*}

\newpage
\begin{figure*}[!htbp] 
	\includegraphics[width=\linewidth]{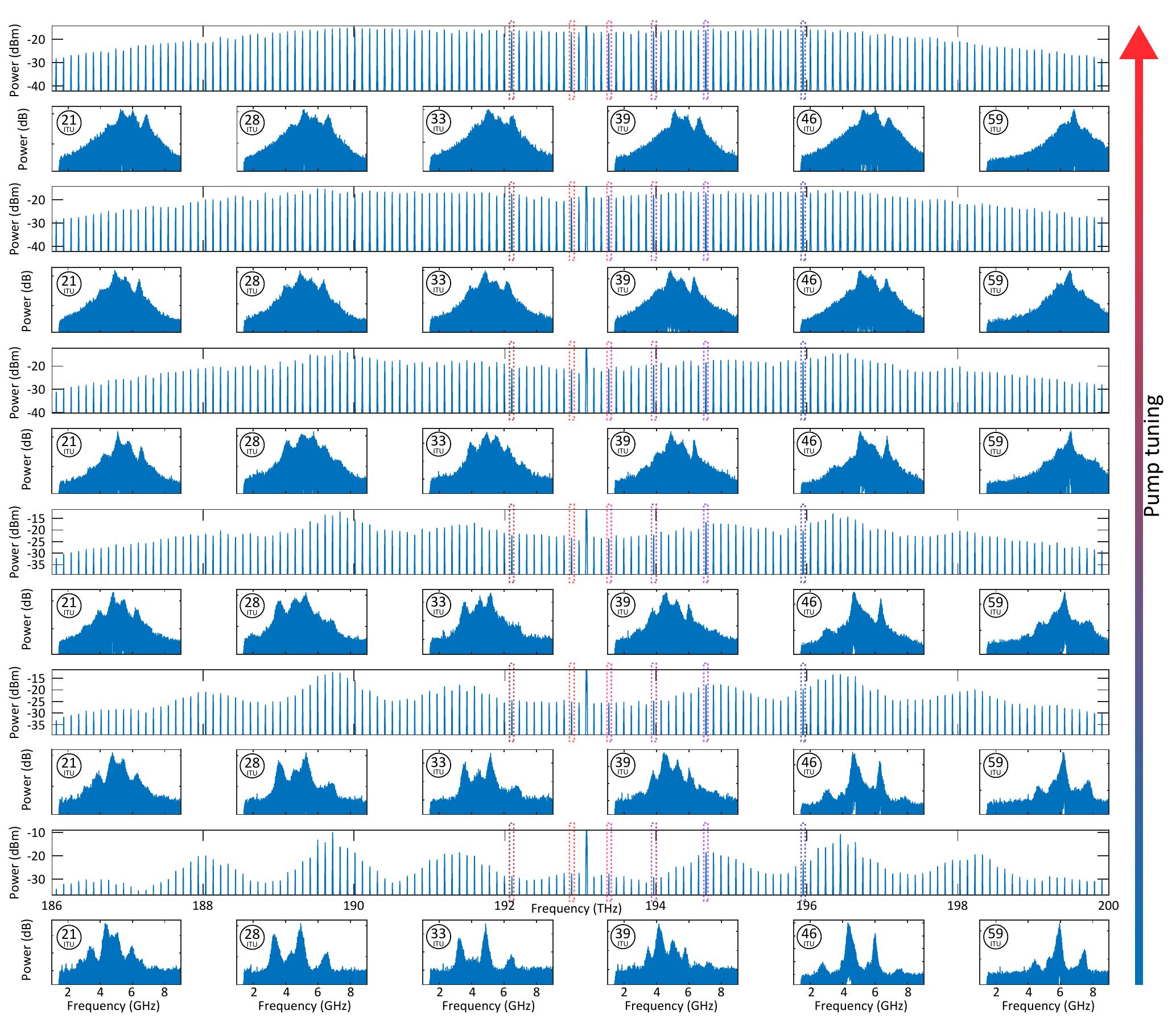}
	\caption{\textbf{Experimental observation of chaotic MI comb state.}
	Experimentally measured optical spectra and heterodyne beatnote power spectral densities (PSD) of selected comb chaotic MI comb lines (ITU: 21, 28, 33, 39, 46, 59) at different laser cavity detuning. The PSD develops from discrete noise features that are derived from the initial frequency comb lines to a broadband continuous spectrum.
	}
	\label{fig_SI_MI_evo_OSA_exp}
\end{figure*}

\newpage
\begin{figure*}[!htbp] 
	\includegraphics[width=\linewidth]{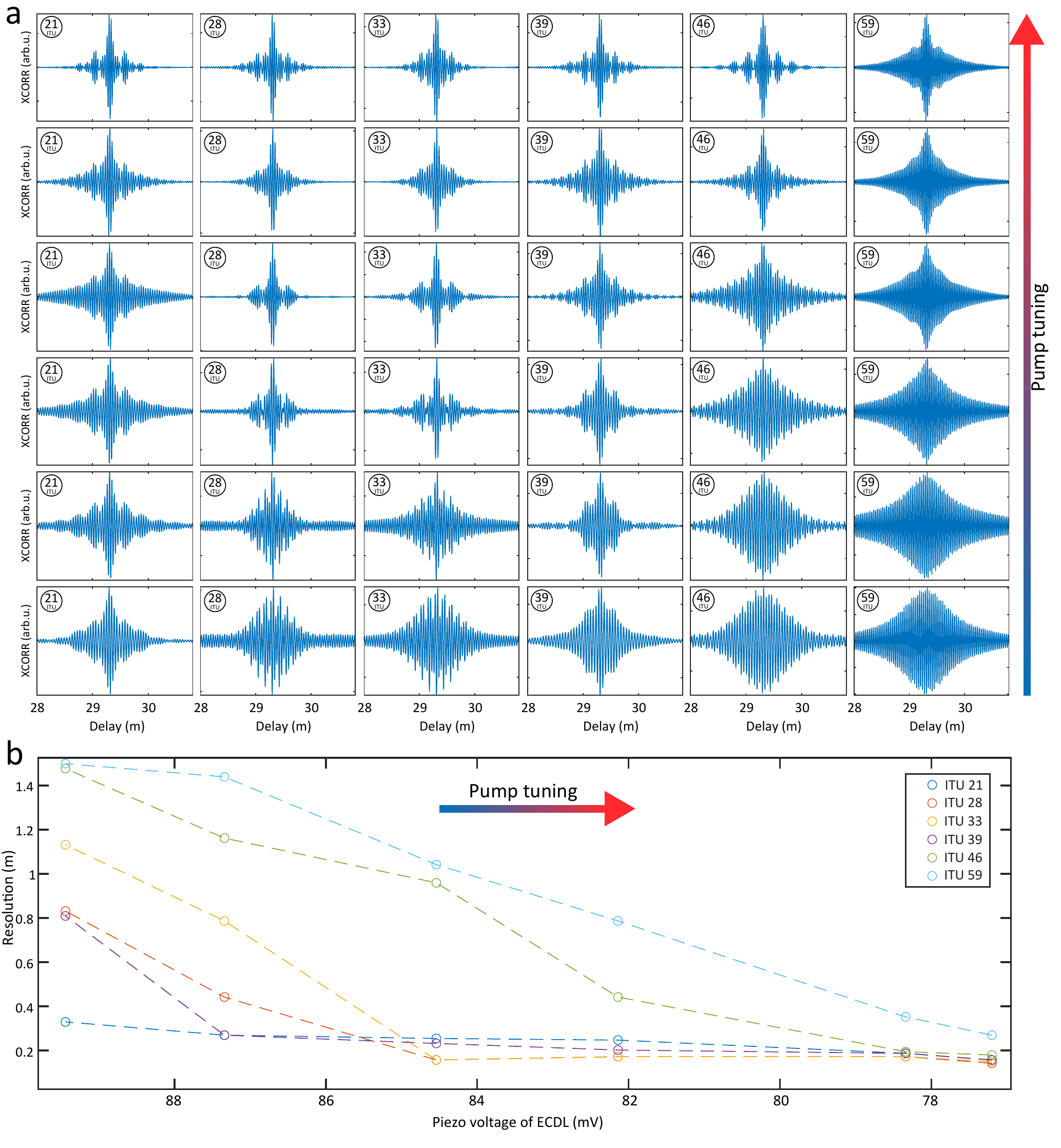}
	\caption{\textbf{Experimental observation of chaotic MI comb state.}
	a)~Auto-correlation of heterodyne beat note signals of chaotic MI comb lines with signal soliton microcomb lines shown in Extended Data Fig. \ref{fig_SI_MI_evo_OSA_exp}. 
	The decreasing width of the autocorrelation peak does indicate increased range resolution for chaotic MI with small detuning.
	b)~Range resolution evolution of selected comb lines inferred as full width at half maximum (FWHM) of auto-correlation traces.
	}
	\label{fig_SI_MI_evo_XCORR_exp}
\end{figure*}

\newpage
\begin{figure*}[!htbp] 
	\includegraphics[width=\linewidth]{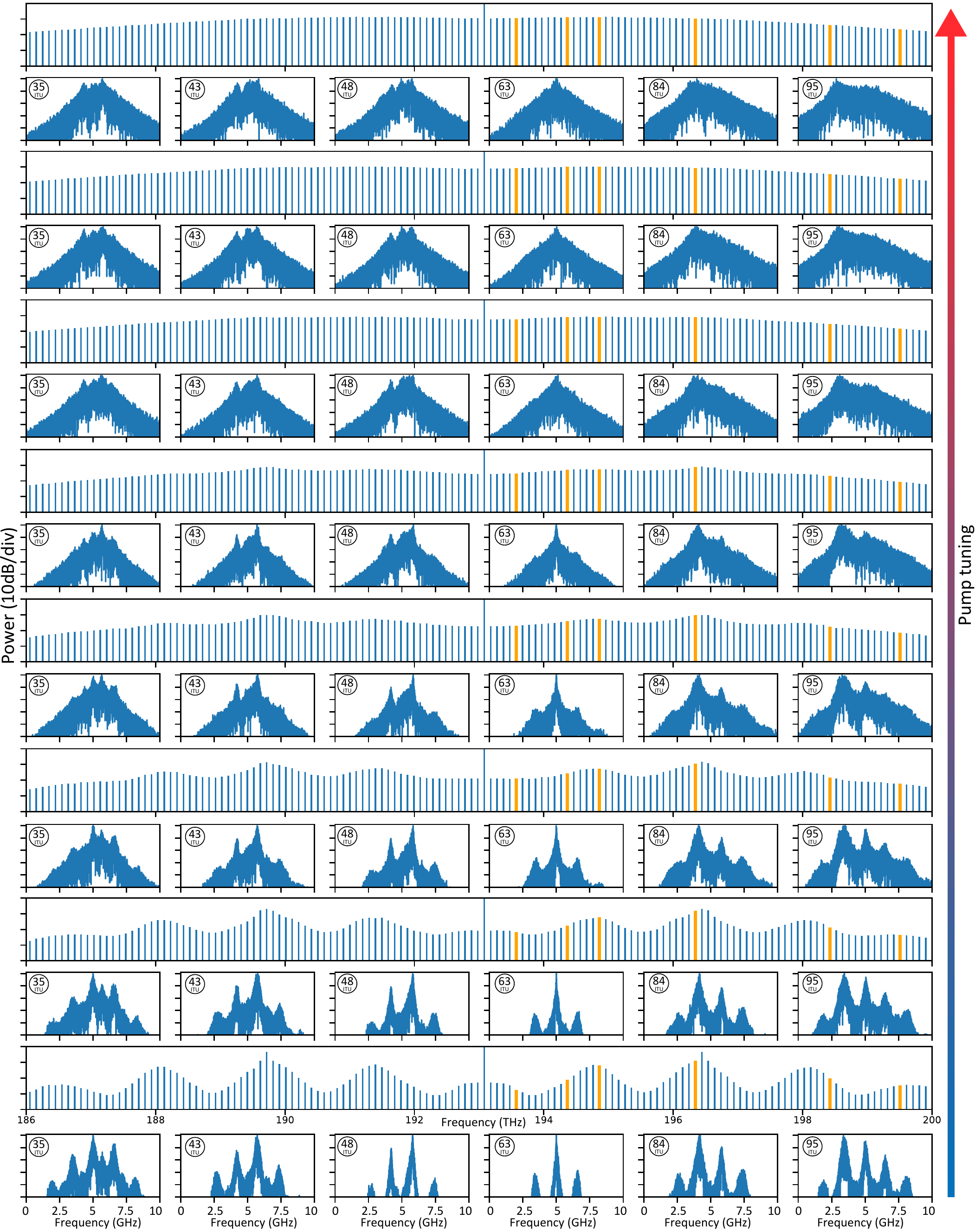}
	\caption{\textbf{Numerical calculations of chaotic MI comb state.}
	Numerically calculated optical spectra (\nth{1},\nth{3},\nth{5} ... row) and heterodyne beatnote power spectral densities (\nth{2},\nth{4},\nth{6}, ... row) of selected comb lines (ITU: 35, 43, 48, 63, 84, 95) for different laser cavity detuning (cf. Ext. Data Fig. \ref{fig_SI_MI_evo_OSA_exp}).
	}
	\label{fig_SI_MI_evo_OSA_sim}
\end{figure*}

\newpage
\begin{figure*}[!htbp] 
	\includegraphics[width=\linewidth]{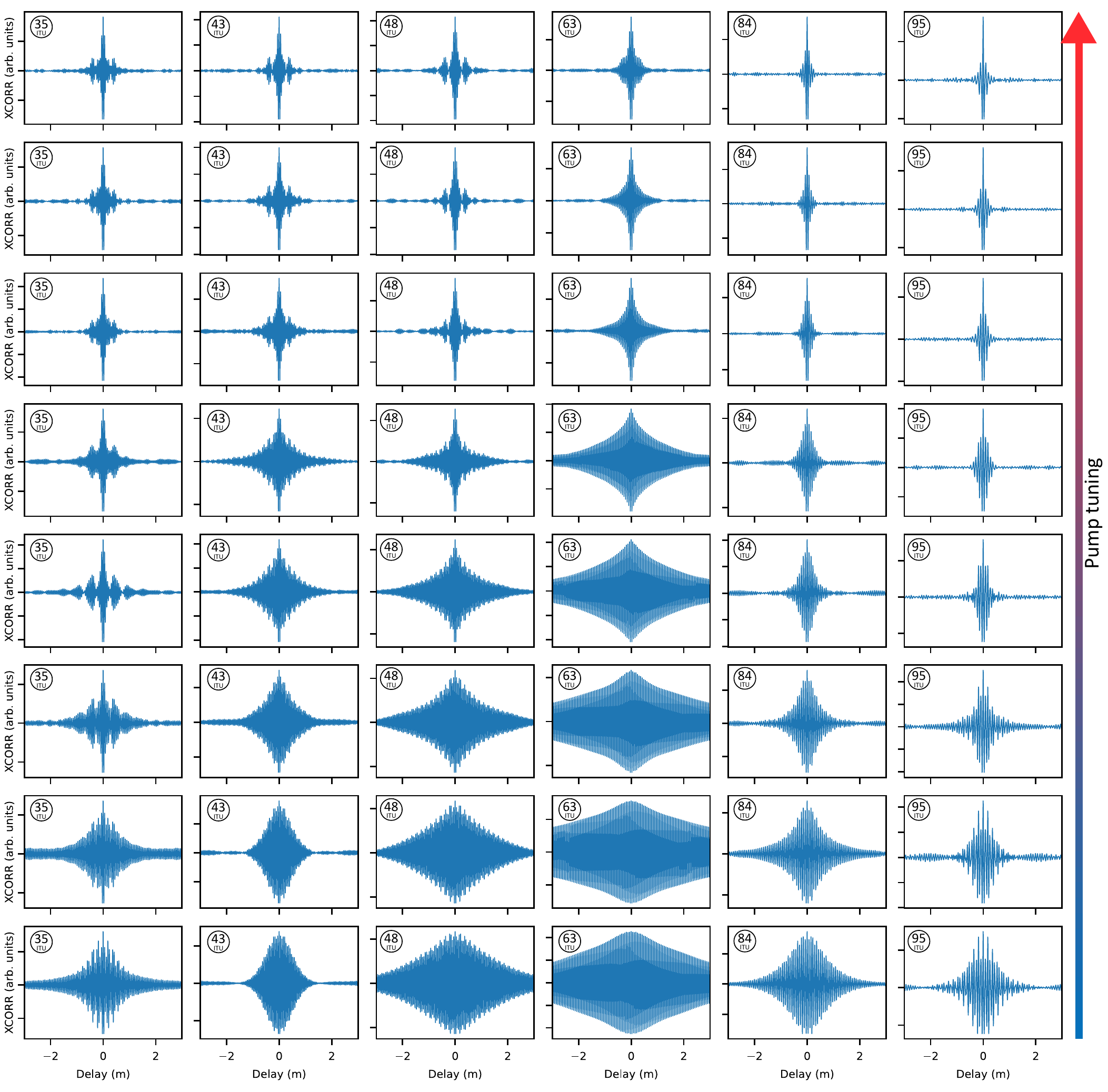}
	\caption{\textbf{Numerical calculations of chaotic MI comb state.}
	a)~Auto-correlation traces of chaotic MI comb lines shown in Extended Data Fig. \ref{fig_SI_MI_evo_OSA_sim} (cf. Ext. Data Fig. \ref{fig_SI_MI_evo_XCORR_exp}).
	}
	\label{fig_SI_MI_evo_XCORR_sim}
\end{figure*}

\newpage
\begin{figure*}[!htbp] 
	\includegraphics[width=\linewidth]{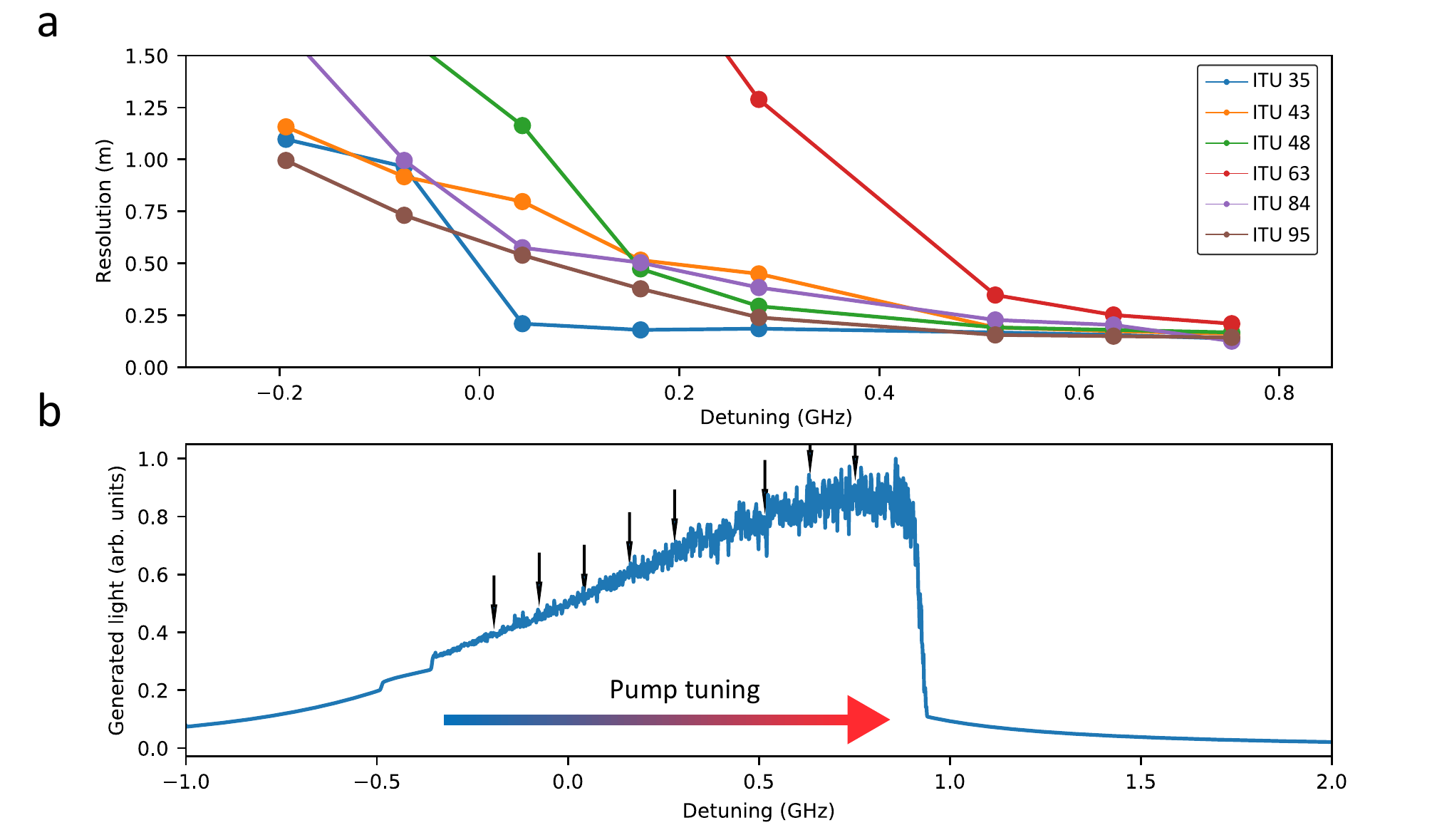}
	\caption{\textbf{Numerical calculations of RMCW LiDAR resolution. }
	a)~Simulation of range resolution evolution of selected comb lines inferred as full width at half maximum (FWHM) of auto-correlation traces depicted in Extended Data Fig. \ref{fig_SI_MI_evo_XCORR_sim}.  (cf. Ext. Data Fig. \ref{fig_SI_MI_evo_XCORR_exp}b).
	b)~Intracavity generated power vs pump detuning. Zero detuning corresponds to the "cold" cavity resonance frequency. Black arrows indicate detuning values used in simulations.
	}
	\label{fig_SI_MI_evo_res_sim}
\end{figure*}

\newpage
\begin{figure*}[!htbp] 
	\includegraphics[width=\linewidth]{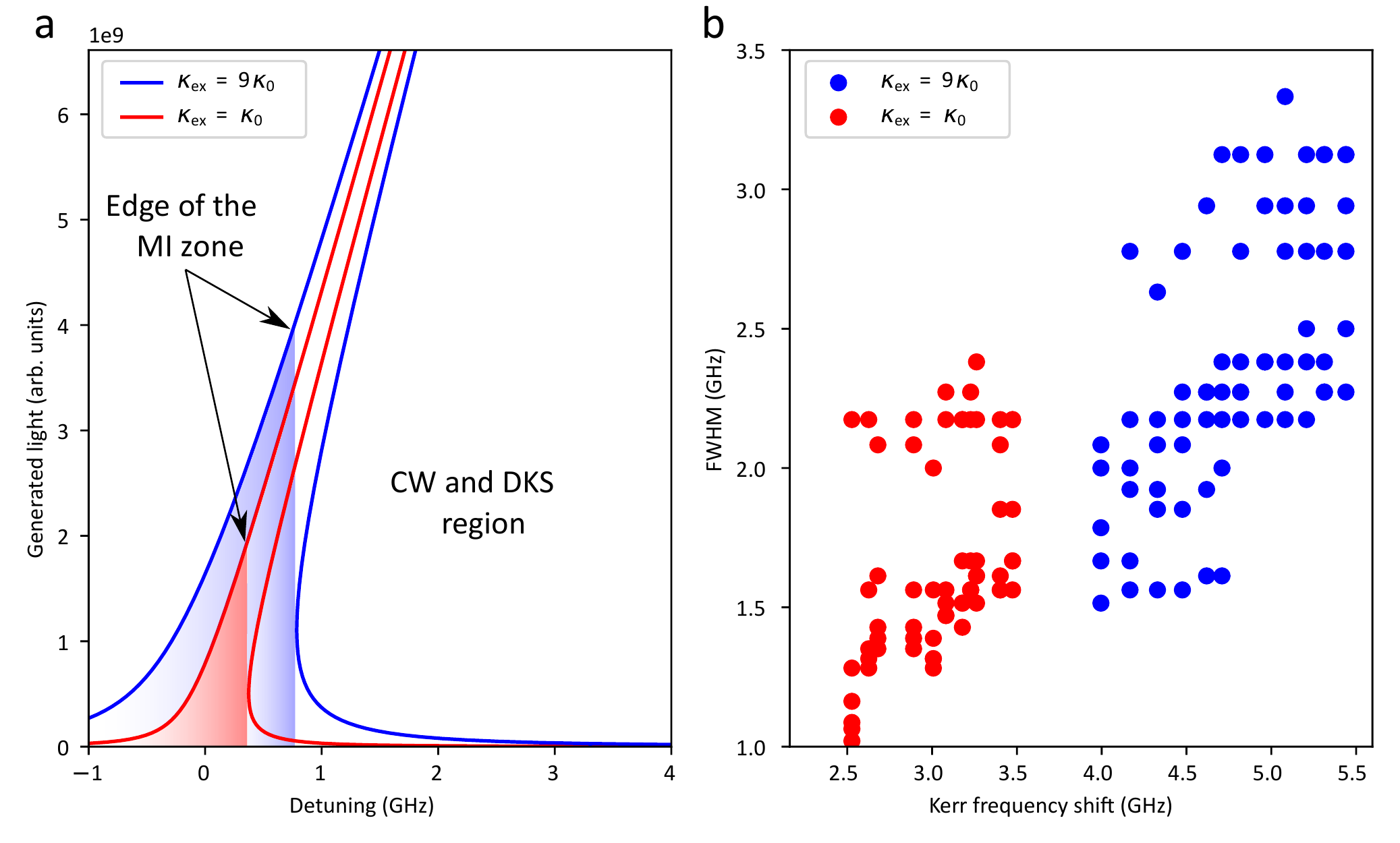}
	\caption{\textbf{Modulation instability noise dependence on microresonator Kerr frequency shift. }
    a)~Tilted resonance for critically coupled $\kappa_\mathrm{ex} = \kappa_0$ and overcoupled $\kappa_\mathrm{ex}=9\kappa_0$ resonanators for $\SI{0.9}{\watt}$ input power.
    b)~Simulated MI state evolution of characteristic noise bandwidth of selected channels as a function of Kerr frequency shift for increasing pump power and fixed pre-soliton switching detuning (very right black arrow on SI Fig. \ref{fig_SI_MI_evo_res_sim}) for critically and overcoupled cases. FWHM is calculated as the speed of light divided by distance resolution inferred from autocorrelation trace.
	}
	\label{fig_SI_FWHM_SPM}
\end{figure*}

\newpage
\begin{figure*}[!htbp] 
	\includegraphics[width=\linewidth]{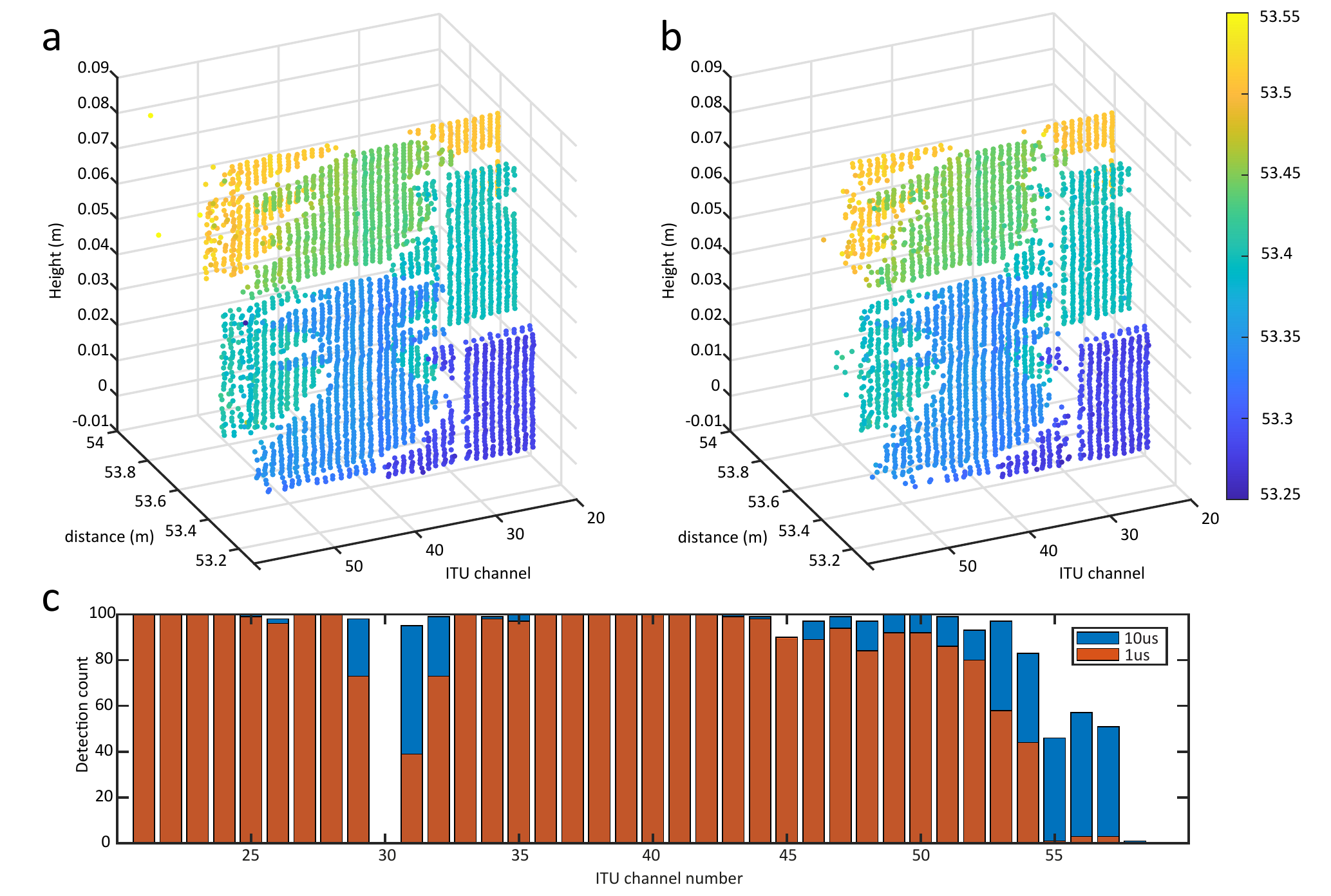}
	\caption{\textbf{Ranging acquisition time variation.}
	a)~Chess imaging with \textbf{10~$\mu$s} pixel acquisition time similar to Fig. \ref{fig_ranging}b.
	b)~Chess imaging with \textbf{1~$\mu$s} pixel acquisition time. Pixels at the figures' edges and corresponding to pump channels have lower detection count.
	c)~Pixel detection count for different acquisition times. In case of not sufficient detection rate RMCW LiDAR can dynamically update its measurement integration time in "real-time" or batch processing.
	}
	\label{fig_SI_chess_1us_10us}
\end{figure*}

\newpage
\begin{figure*}[!htbp] 
	\includegraphics[width=\linewidth]{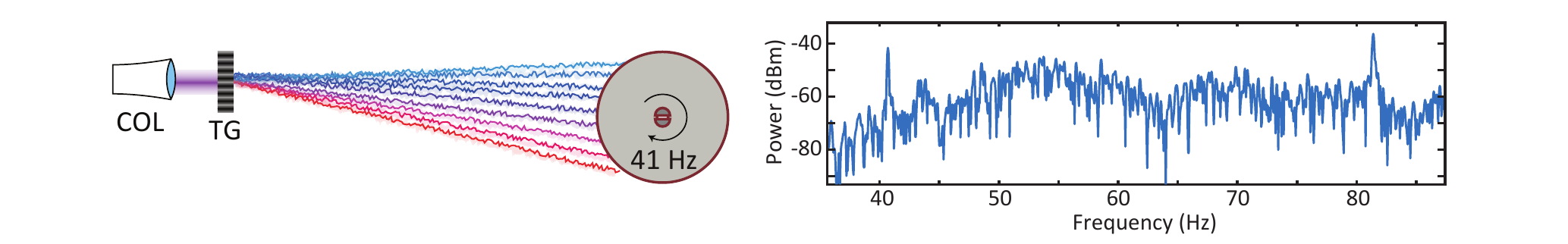}
	\caption{\textbf{Flywheel audio power spectral density.}
	(Left) Flywheel imaging setup. (Right) Power spectral density of the audio trace of the rotating flywheel recorded on a cell-phone microphone. The 41~Hz peak corresponds to the fundamental harmonic of the mechanical motion. Higher harmonics are observed due to mechanical vibrations.
	}
	\label{fig_SI_wheel_PSD}
\end{figure*}

\newpage
\begin{figure*}[!htbp] 
	\includegraphics[width=\linewidth]{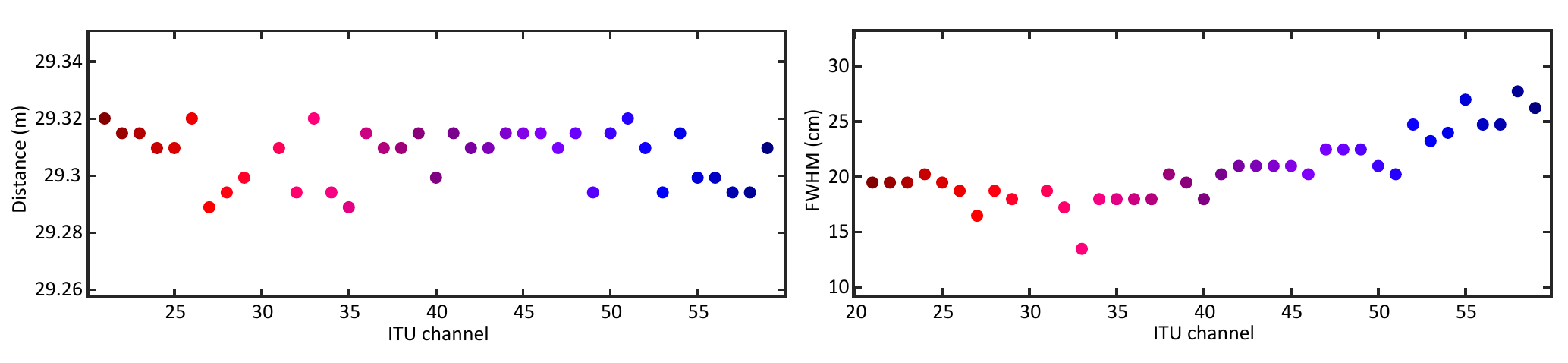}
	\caption{\textbf{Direct detection RMCW LiDAR.}
	(Left) Fiber delay measurement via direct detection of MI based RMCW LiDAR. (Right) FWHM of cross-correlation traces corresponding to distance resolution of direct detection  RMCW LiDAR. The values of channel dependent resolution correspond to the case of coherent RMCW LiDAR where the signals were digitally high-pass filtered with 550~MHz pass frequency.
	}
	\label{fig_SI_DD}
\end{figure*}

\newpage
\begin{figure*}[!htbp] 
	\includegraphics[width=\linewidth]{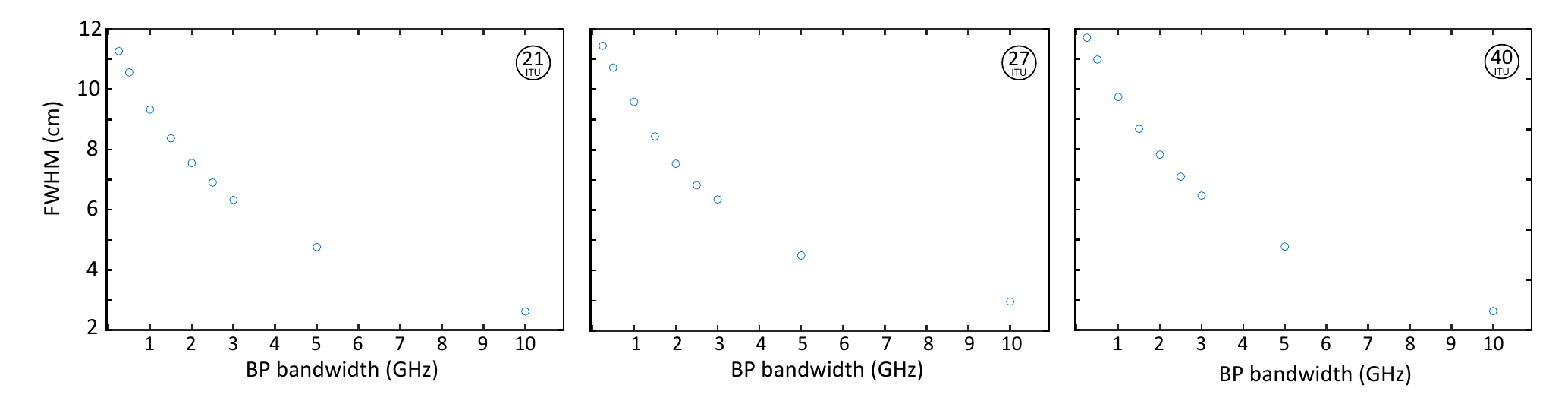}
	\caption{\textbf{ASE based LiDAR.}
	Broadband amplification spontaneous emission noise from erbium doped fiber amplifier is used to determine achievable distance resolution. Figures depict FWHM or distance resolution versus digital band-pass filter bandwidth for 21, 27, and 40 ITU channels correspondingly.
	}
	\label{fig_SI_ASE}
\end{figure*}

\newpage
\begin{figure*}[!htbp] 
	\includegraphics[width=\linewidth]{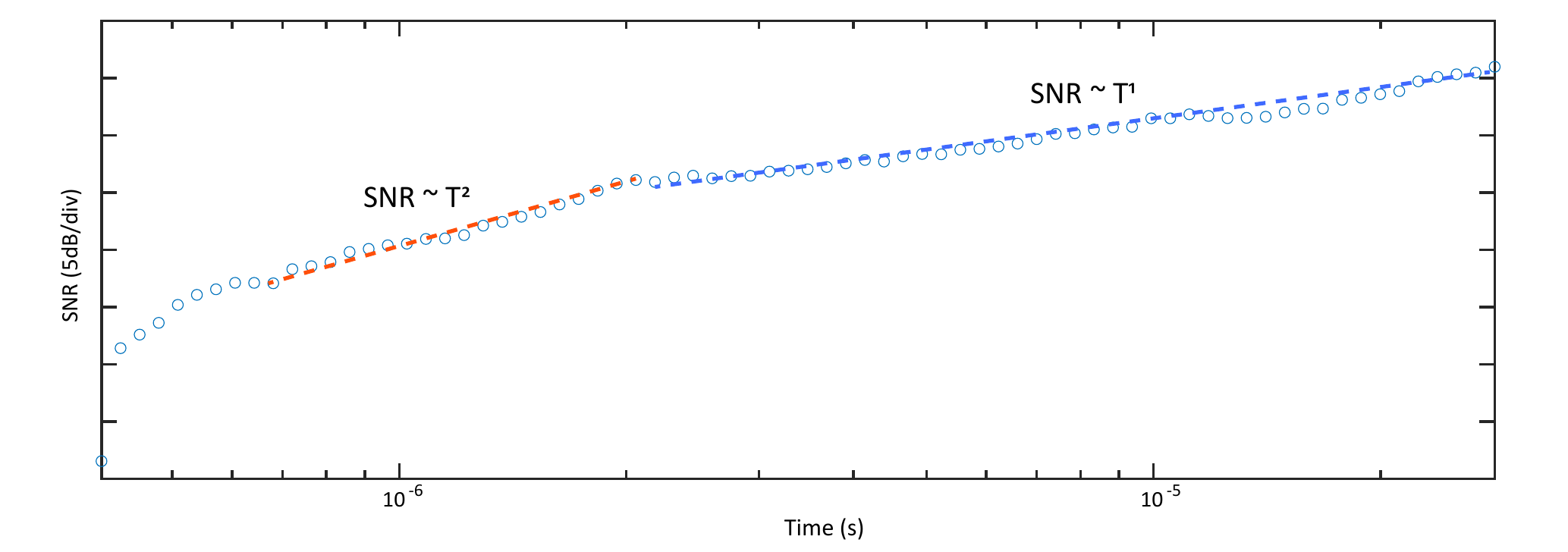}
	\caption{\textbf{SNR of the cross-correlation trace time dependence.}
	 ITU 21 channel SNR versus measurement time dependence for a free-space measurement. Blue dashed line highlights a linear dependence in accordance with formula \ref{eq:SNR_xcorr_final} for measurement times more than delay time equal to $\sim$~350~ns.
	}
	\label{fig_SI_SNR_time}
\end{figure*}

\newpage
\begin{figure*}[!htbp] 
	\includegraphics[width=\linewidth]{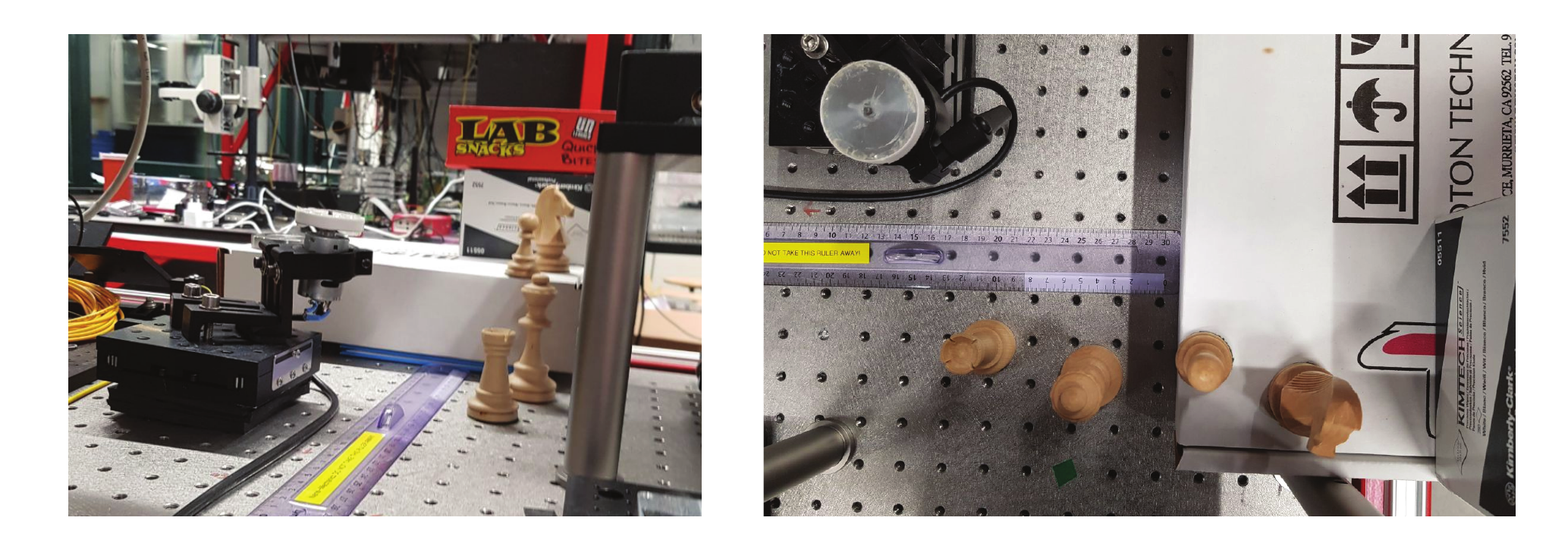}
	\caption{\textbf{Imaging scene.}
	Photographs showing chess figures and a flywheel used in 3D imaging and velocimetry.
	}
	\label{fig_SI_scene}
\end{figure*}

\end{document}